\newcommand{\be}{\begin{equation}}
\newcommand{\ee}{\end{equation}}

\documentclass[]{napolitano}
\usepackage{epsfig}
\begin{document}
%
\title{Planetary nebulae as mass tracers of their parent galaxies: biases
in the estimate of the kinematical quantities}
%
%
%
\author{Nicola R. Napolitano \inst{1,2}
\and Magda Arnaboldi \inst{1}
\and Kenneth C. Freeman \inst{3}
\and Massimo Capaccioli \inst{1,2}
}
%
%
\institute{Osservatorio Astronomico di Capodimonte, via Moiariello 16,
I-80131 Napoli, Italy
\and
              Dipartimento di Scienze Fisiche, Universit\`a\ Federico II 
              di Napoli,
              Complesso Monte S.Angelo, via Cintia, 80126, Napoli, Italy
\and
RSAA, Mount Stromlo Observatory, Australian National University, Weston 
Creek P.O., ACT 2611 }
\date{Received:... ; accepted:...}
\titlerunning{Extragalactic planetary nebulae as mass tracers}
\authorrunning{Napolitano et al.}

\abstract{
Multi-object and multi-fiber spectrographs on 4 and 8 meter telescopes
make it possible to use extragalactic planetary nebulae (PNe) in the
outer halos of early-type galaxies as kinematical tracers, where
classical techniques based on integrated stellar light fail. Until
now, published PNe radial velocity samples are small, with only a few
tens of radial velocity measurements (except for a few cases like NGC
5128 or M31), causing uncertainties in the mass and angular momentum
estimates based on these data. To quantify these uncertainties, we
have made equilibrium models for spherical galaxies, with and without
dark matter, and via Montecarlo simulations we produce radial velocity
samples with different sizes. We then apply, to these discrete radial
velocity fields, the same standard kinematical analysis as it is commonly
done with small samples of observed PNe radial velocities. By
comparison of the inferred quantities with those computed from the
analytical model, we test for systematic biases and establish a robust
procedure to infer the angular momentum distribution and radial
velocity dispersion profiles from such samples.
}
\maketitle

\section{Introduction\label{Intro}} The dynamics of the outer regions of
early-type galaxies have been studied by means of test particles like globular
clusters (GCs) and planetary nebulae (PNe). GCs have been successfully used to
probe the gravitational potential in giant ellipticals (M87: Mould et al. \cite{moul};
Cohen \cite{cohen}; NGC5128: Sharples \cite{sharp}; Harris et al. \cite{harr}; NGC1399: Grillmair et
al. \cite{grill}; Minniti et al. \cite{minn}; Kissler-Patig et al. \cite{kpal}), but their number
density and angular momentum distributions often turn out to be different from
those of the stellar population in the outer galaxy halos. 

PNe are a population of dying stars, whose outer envelope re-emits more
than 15\% of the energy emitted in the UV by the internal star in the
[OIII] green line at 5007 \AA~ (Dopita et al. \cite{dopi}) and therefore they
can be readily detected in distant galaxies. The observational
evidences indicate that the number density of PNe, unlike GCs, is
proportional to the underlying stellar light (Ciardullo et al. \cite{ciar89};
Ciardullo et al. \cite{ciar91}; McMillan et al. \cite{mcmil}; Ford et al. \cite{ford}) and they
share the angular momentum distribution of the stars (Arnaboldi et al. \cite{arn1}; Hui et al. \cite{hui}; Arnaboldi et al. \cite{arn2}, \cite{arn3}).
First attempts to use PNe as kinematical tracers in nearby galactic
systems date back to 1986 (Nolthenius \& Ford \cite{nolth}) and studies of
early-type systems within a distance of 10 Mpc rapidly followed
(Ciardullo et al. \cite{ciar93}; Tremblay et al. \cite{trem}; Hui et al. \cite{hui}). Since
1993 new observing techniques (eg Arnaboldi et al. \cite{arn1}) allowed
measurements of radial velocities of PNe in the outer regions of giant
early-type galaxies situated at distances larger than 10 Mpc. These
studies, mostly based on samples of only a few tens of PNe, show that
the outer PNe typically have faster systemic rotation than GCs
(Grillmair et al. \cite{grill}; Arnaboldi et al. \cite{arn1}; Hui et al. \cite{hui}; Arnaboldi et al. \cite{arn3}).

So far these small samples were analysed by adopting simple
three-parameter functions for the underlying projected rotation field,
but no detailed studies have yet been reported to test for biases in the
estimated rotation and velocity dispersion introduced by these adopted
(parametric) functions. Non-parametric analyses can in principle avoid
this problem. In practice, however, for the small data samples of
interest here, the associated inherent smoothing effectively drives the
non-parametric analysis towards one of the commonly used simple
parametric forms (like solid body rotation: see for example Arnaboldi
et al. \cite{arn3}).

In the large telescope era, much larger samples of PNe velocities will
become available for galaxies out to distances of about 20 Mpc. The
question of biases induced by adopting simple parametric forms for the
rotation field remains relevant, because studies of galaxies at larger
distances will still be limited to small sample of radial velocities.
It will be important to know how to compare the results for nearby giant 
ellipticals, based on large samples of PNe velocities, with those for
more distant objects, based on smaller samples.

In this paper we build simple equilibrium models with and without dark
matter (DM) for spherical early-type galaxies (Sect.~\ref{din}). Via
Montecarlo simulations, we extract samples of tracers of various sizes
(Sect.~\ref{SDRF}). We then adopt simple forms for the mean rotation
field and derive the parameters for the projected 2-D rotation velocity
field from these samples (Sect. 4 and 5). We also derive estimates of
the precision and biases for the mean rotation and velocity dispersion
(Sect.~\ref{res} and 7). The main goal of this work is to determine
the minimum sample needed to derive reliable estimates of the
kinematics of the host system via these parametric fits, and to define
a robust approach to derive observables from small radial velocity
samples. Discussion and conclusions are drawn in Sect.~\ref{conclu}.

\section{Analytical spherical systems in equilibrium \label{din}}
In our modeling we wish to represent the characteristics of the observed
PNe distribution in early-type galaxies. In these systems, the observed
PNe number density follows the stellar light distribution (Ciardullo et al. 
\cite{ciar89}; Ciardullo et al. \cite{ciar91}; McMillan et al. \cite{mcmil}; Ford et al. \cite{ford}), except
in the bright inner regions where the PNe counts become incomplete. As the 
surface brightness of the stellar continuum background increases towards 
small radii, the 5007 \AA~[OIII] emission from the PNe is more difficult 
to detect and the apparent ratio of PNe to luminosity decreases. In Appendix, we compute the value of the limiting radius $R_\mathrm{lim}$ for which at
large radii the PNe sample is complete. For the surface brightness profile
of a typical E galaxy in Virgo Cluster, we find that $R_\mathrm{lim} = 0.7
\times R_\mathrm{e}$, where $R_\mathrm{e}$ is the effective radius.

\subsection{Systems without dark matter}
Assuming constant mass to light ratio (M/L), the analytic Hernquist (\cite{hern})
model is a good approximation to a system whose surface brightness 
distribution follows the de Vaucouleurs law (\cite{devau}). 
The luminous mass density is given by:
\be
\rho(r)=C_\mathrm{l} \frac{M_\mathrm{l} a}{2\pi}\frac{1}{r(r+a)^{3}}\mbox{,}
\label{rher}
\ee
where $M_\mathrm{l}$ is the total luminous mass, $a$ is a distance scale 
($R_\mathrm{e}=1.8153~a$) and $C_\mathrm{l}$ is a normalization constant. We consider systems
truncated at $R_\mathrm{max}=18a$ (i.e. $\sim 10 R_\mathrm{e}$). For $a=1$ and $M_\mathrm{l}=1$, 
the normalization constant $C_\mathrm{l}=1.114$. Writing $r$ for $r/a$, 
we define a dimensionless density distribution
\be
\tilde{\rho} (r)=\left\{
\begin{array}{lll}
\frac{C_\mathrm{l}}{2\pi}\frac{1}{r(r+1)^3} & & r \leq 18 \\
0 & & r>18
\end{array}
\right. .
\label{henor}
\ee
The cumulative mass distribution is then
\be
M_\mathrm{l}(r)=4\pi \int_0 ^r x^2 \rho(x)dx.
\label{mascu}
\ee
~\\
The test particles are extracted from this distribution, taking into account
the incompleteness effects at radii $\le R_\mathrm{lim}$. We then consider the
kinematics of PNe within $10a = 5.5 R_\mathrm{e}$, corresponding to the typical
radial coverage of the observed PNe samples in real galaxies.

\subsubsection{Systems with dark matter}
\begin{figure}
\centering
\epsfig{figure=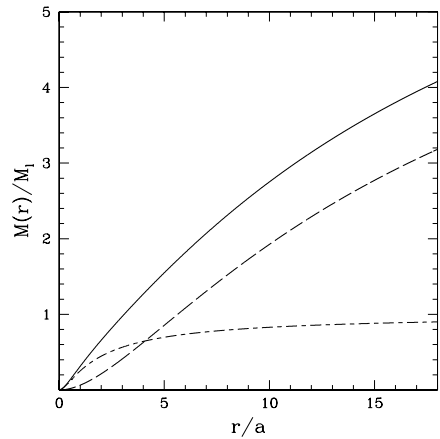,width=6cm,height=6cm}
\caption{\footnotesize Mass distributions for the Hernquist model:
luminous mass matter (dash-dotted line), dark halo mass (dashed line) and
the total mass (solid line).}
\label{masses}
\end{figure}

In constructing our equilibrium models, we also consider systems with
an additional mass contribution to the stellar mass density coming from
a dark halo. Furthermore we model the dark halo with
a Hernquist mass distribution, with scale length large enough so that it
roughly mimics a $\rho \sim r^{-2}$ halo in the region of interest ($r
< 10$). We write the dark matter density as
\be
\rho_\mathrm{d}(r)= \frac{M_\mathrm{d} d}{2\pi}\frac{1}{r(r+d)^{3}}
\label{rhdark}
\ee
and adopt $d=10a$ and $M_\mathrm{d}=7.7 M_\mathrm{l}$ (in agreement with the mass distribution of NGC5128, Hui et al. \cite{hui}).
The cumulative mass distribution is defined as in Eq.~(\ref{mascu}), and
the potential is derived from the total mass
\be
M_\mathrm{t}(r)=M_\mathrm{l}(r)+M_\mathrm{d}(r).
\label{mast}
\ee
The $M_\mathrm{l}(r)$, $M_\mathrm{d}(r)$ and $M_\mathrm{t}(r)$ distribution are shown in
Fig.~\ref{masses}.

\subsection{Intrinsic and projected kinematics \label{kin}}
We consider non-rotating and rotating equilibrium spherical systems,
with the velocity dispersion components given by the Jeans
equations\footnote{The possibility that spherical systems can rotate
was investigated by Lynden-Bell (\cite{lynbel}): here we consider such systems
in order to use simple density and gravitational potential
functions.}.\\ For non-rotating systems with no dark matter, we
assume that mass follows light and adopt an isotropic velocity
dispersion to solve the radial Jeans equation

\be
\frac{ d (\rho\sigma ^2 )}{d r}=-G\frac{M_\mathrm{l}(r)\rho (r)}{r^2 }.
\label{istnr}
\ee
The solution in this simple case is 
\be
\begin{array}{rl}
\sigma^2(r)=&7.1 10^{-7}r(1.+r)^3\{-0.13-\\
	&[130321(1+r)^{-4} (25+52r+42r^2+ 12r^3\\
            &+12(1+r)^4\log(r)-12(1+r)^4\log(1+r))]\}	    
\end{array}
\label{nrsol}
\ee
where $\sigma(r)$ is given in units of $v_\mathrm{g}=\sqrt{GM_\mathrm{l}/a}$$^($\footnote{If we take 
$M_\mathrm{l}=3.5\times 10^{11} M_\odot$ and $a=2$ kpc, we have $v_\mathrm{g}=870~$km/s, 
so the velocities in our simulations are of the order of $0.3v_\mathrm{g}=260$ km/s.}$^)$.\\

We should also investigate rotating systems because early-type galaxies
appear to show fast rotation in their outer parts (Hui et al. \cite{hui};
Arnaboldi et al. \cite{arn2}, \cite{arn3}). What is the most appropriate form for the
rotation law ? The only system for which a mean stellar rotation law
has been derived reliably out to large radii is Centaurus A. Here, Hui
et al. (\cite{hui}) adopt the functional form
\be
v_\mathrm{rot}(R)=\frac{V_\mathrm{max} R}{\sqrt{R^2 +R_\mathrm{v} ^2 }},
\label{rotin} 
\ee
where $R$ is the distance from the galactic center and $R_\mathrm{v}$ a scale distance,
which reproduces the rotation of NGC5128 in its equatorial plane. 
We assume that our model systems have the same underlying rotation law,
and adopt a cylindrical rotational structure with the same functional form, 
where $R$ is now the radial coordinate in a cylindrical ($R$, $\varphi$, 
$z$) system.

In a conventional cylindrical coordinate system, ($R$, $\varphi$, $z$),
the Jeans 
equations become 
\be
\frac{\partial \rho \overline{v_R^2} }{\partial R}+ \rho \left(\frac{\overline{v_R^2}-\overline{v_{\varphi}^2}}{R}+
\frac{\partial \Phi}{\partial R}\right)
=0 \mbox{ and}
\label{JE1}
\ee
\be
\frac{\partial \rho \overline{v_z^2}}{\partial z} + \rho \frac{\partial \Phi}{\partial z}=0,
\ee
where $\rho$ and $\Phi$ are functions of ($R$, $z$).

If we substitute into equation (\ref{JE1}) a given cylindrical rotation law,
and assume an isotropic velocity dispersion, then this system is over-determined.
If we now relax the condition of isotropy and assume that the principal
axes of the velocity ellipsoid lie parallel to the ($R$, $z$) axes and
$\sigma_R^2=\sigma_{\varphi}^2 \neq \sigma_z^2$, the system of
equations can be written as
\be
\begin{array}{ll}
\frac{\partial\rho(R,z)\sigma_R^2(R,z)}{\partial R} - \frac{\rho(R,z)}{R} v_\mathrm{rot}^2(R)
=& \\
 & \\
~~~~~~~~~~~~~~~~~~~~~=- \rho(R,z) \frac{\partial \Phi(R,z)}{\partial R} &\mbox{ and}
\label{axJ1}
\end{array}
\ee
\be
{\footnotesize \frac{\partial \rho(R,z) \sigma_z^2(R,z)}{\partial z} 
= - \rho(R,z) \frac{\partial \Phi(R,z)}{\partial z}}.
\label{axJ2}
\ee
These two equations can be solved independently by simple inversion as
in Binney \& Tremaine (\cite{binn})\footnote{They solve the second equation in
the isotropic case (page 120).}. We believe this to be a reasonable
approach for our purpose. It describes an equilibrium system with a
spherical potential and a given cylindrical rotation law of known
functional form\footnote{Our assumptions on the alignment and
anisotropy of the velocity dispersion tensor were motivated by the need
for a simple approach to the solution of the Jeans equations. Our 
choice is also physically motivated by Arnold (\cite{arnold}) who obtained an
axisymmetric solution which is quasi-cylindrically aligned and where
$\sigma_R^2 \sim \sigma_{\varphi}^2 \neq \sigma_z^2$ (model $vi$). An
advantage of this approach is that an analytical solution can be
obtained and used during our simulations.}.

\begin{figure*}
\centering
\hspace{-1cm}
\epsfig{figure=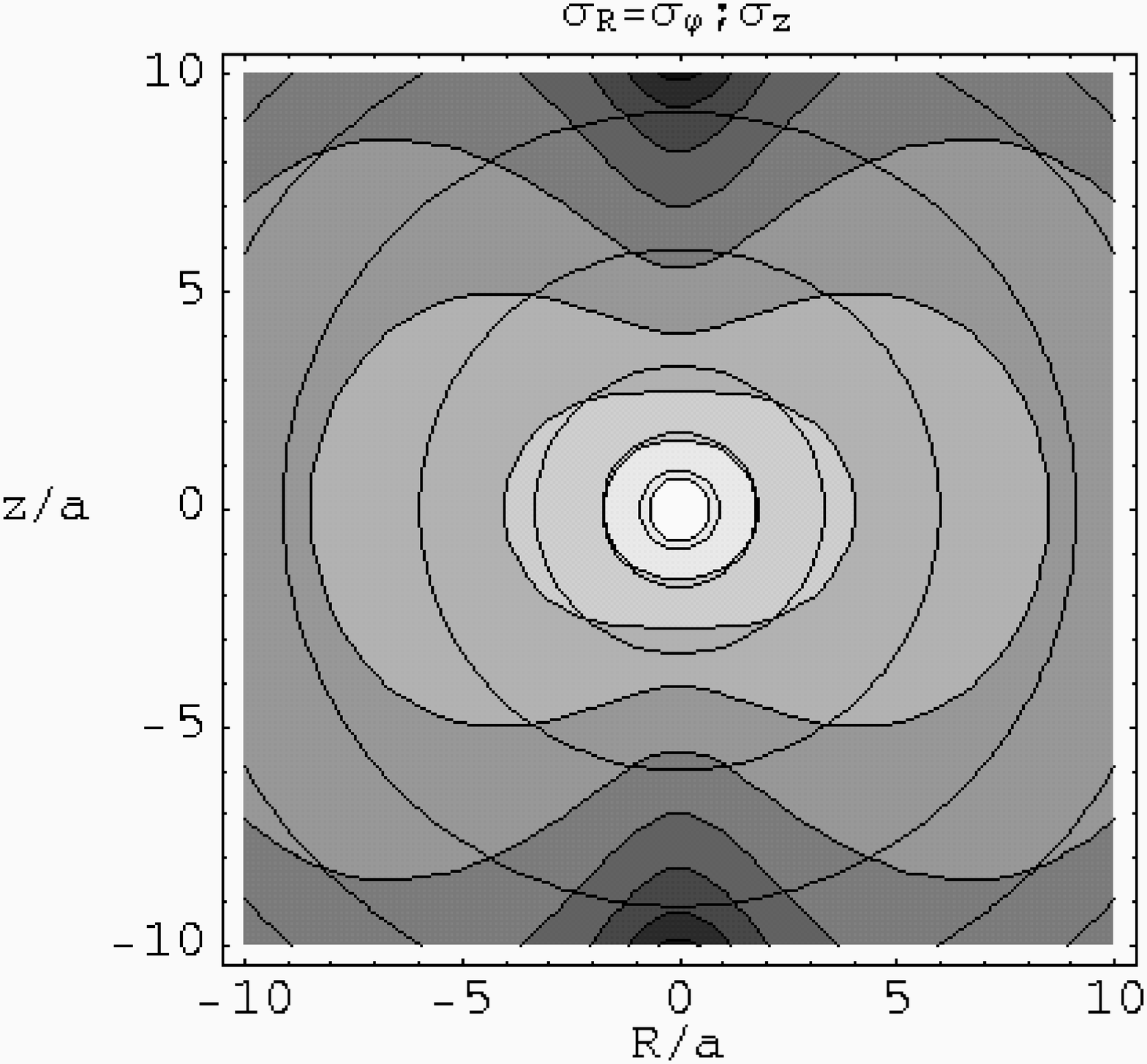,width=6.5cm,height=6cm}
\epsfig{figure=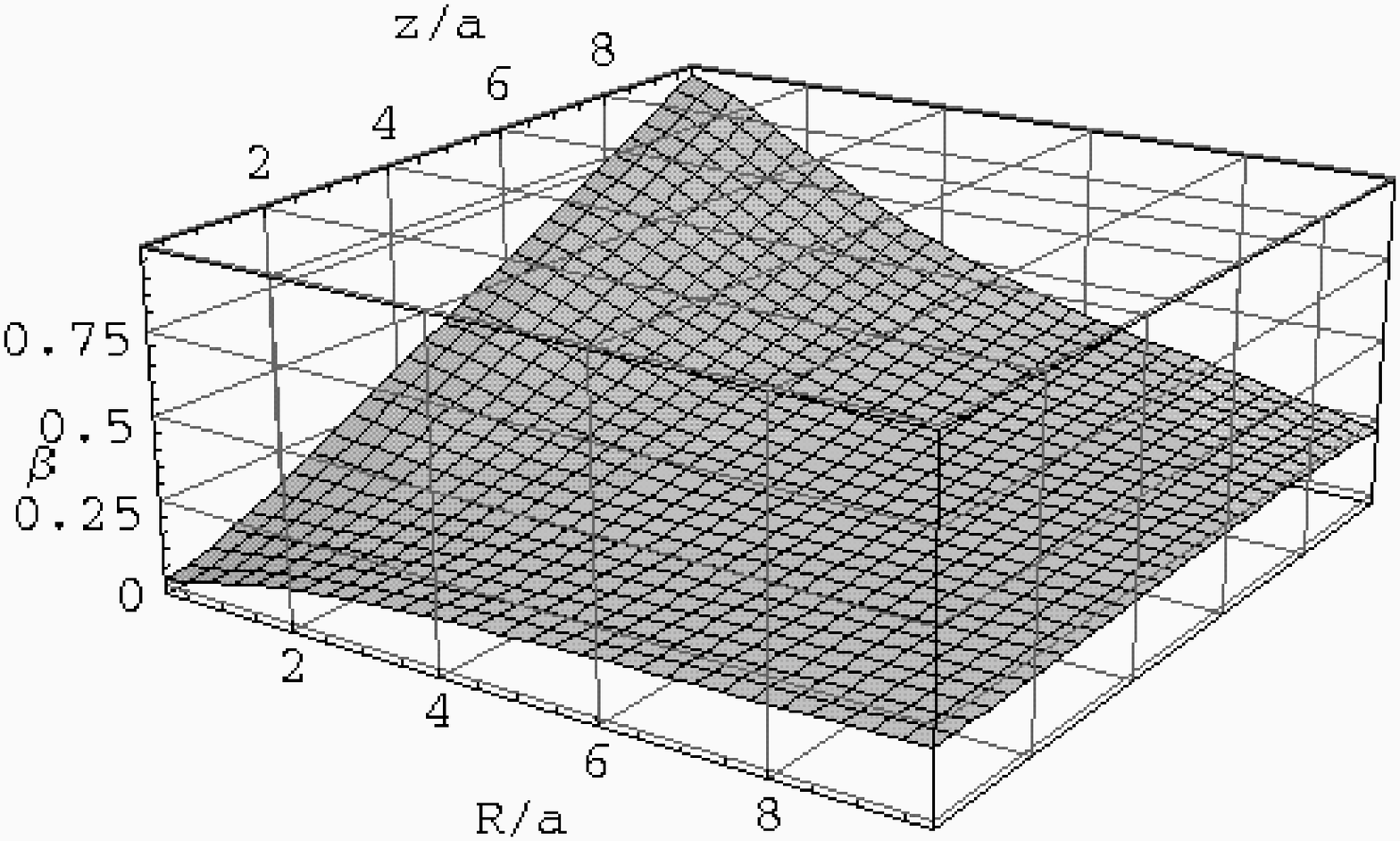,width=7cm,height=5cm}
\caption{\footnotesize $R$ and $z$ dependence of the velocity dispersion 
ellipsoid (left) and anisotropy parameter $\beta=1-\sigma_z^2/\sigma_R^2$ (right) for the rotating model with DM and 
$V_\mathrm{max}=0.23v_\mathrm{g}$. In the left panel the contours of the 
$\sigma_R(=\sigma_{\varphi})$ component 
($\sigma_R/v_\mathrm{g}$=0.325 to 0.125 with step=0.025 from inside) are 
plotted with different grey scales; the circular lines are the 
corresponding $\sigma_z$ contours.}
\label{cont}
\end{figure*}

Physically plausible solutions require that the components of 
the velocity ellipsoid are all positive in the volume occupied by the system. 
As we will see later, only certain values of $V_\mathrm{max} $ in Eq.~(\ref{rotin}) 
allow physical solutions in the whole volume occupied by the system. 
In Fig.~\ref{cont} we show the intrinsic kinematics for $V_\mathrm{max} =0.23v_\mathrm{g}$ 
and $R_\mathrm{v}=2a$.

\subsection{Geometry and projected kinematics}
As the galaxy intrinsic coordinate system we adopt a Cartesian coordinate 
system where the X-axis coincides with the direction of the line-of-sight 
and the Y-Z plane is the sky plane; the positive direction of the X-axis 
is towards the observer, while $V_\mathrm{rad}=-v_x$ is the radial velocity. 
In case of inclined systems, we adopt the conventional Euler matrix 
to identify the coordinate system X$'$Y$'$Z$'$, where the X$'$-axis 
is coincident with line-of-sight and the Y$'$-Z$'$ plane is
the Sky plane. The three Euler angles are $\phi$, $\theta$, and $\psi$ 
(Goldstein, H. \cite{gold}): when  $\phi=\theta=\psi=0$, i.e. the edge-on case, 
the two coordinate systems XYZ and X$'$Y$'$Z$'$ are coincident.\\
The projected rotation and velocity dispersion profiles
are computed from the intrinsic profiles as follow. The projected
moments of the radial velocities along the line-of-sight are given by
the Abel integral
\be
\langle V_\mathrm{los}^{(n)}\rangle=\frac{2}{I(\tau)}\int_{R}^{\infty} \overline{V_\mathrm{rad}^n}
\frac{\rho r}{\sqrt{r^2-\tau^2}}dr
\label{vlos}
\ee
where $\tau$ is the distance from the galactic center along a fixed direction on the sky plane, 
$I(\tau)$ is the projected density, and 
$\overline{V_\mathrm{rad}^n}$ is the $n$-th moment of the radial velocity in 
each volume element
along the line-of-sight. When we consider inclined systems, we have $V_\mathrm{rad}=-v_x'=(a_{11}v_x+a_{12}v_y+a_{13}v_z)$ 
in Eq.~(\ref{vlos}) where $a_{ij}$ are the elements of the Euler matrix.
$V_\mathrm{rad}$ can be expressed in terms of the cylindrical velocity components:
\begin{eqnarray}
\label{vcilc}
v_x & = & v_R \cos \varphi~-~v_{\varphi} \sin \varphi \nonumber\\
v_y & = & v_R \sin \varphi~+~v_{\varphi} \cos \varphi \\ 
v_z & = & v_z \nonumber
\end{eqnarray}
where $\varphi$ is the conventional azimuthal angle in spherical and cylindrical coordinate systems.
Then, $V_\mathrm{rad}=-(c_1 v_R+c_2 v_{\varphi}+c_3 v_z)$ where
\begin{eqnarray}
\label{vradcf}
c_1 & = & a_{11}\cos \varphi + a_{12} \sin \varphi \nonumber\\
c_2 & = & -a_{11}\sin \varphi + a_{12} \cos \varphi \\ 
c_2 & = & a_{13}. \nonumber
\end{eqnarray}
When Eqs. (\ref{vcilc}) and (\ref{vradcf}) are substituted in Eq.~(\ref{vlos}),
we derive the projected kinematics along the line-of-sight.\\
The rotation curve expected from the model at equilibrium is given by 
$\langle V_\mathrm{los} (\tau)\rangle$, and the velocity dispersion profile 
is given by $\sigma_\mathrm{V}^2=\langle V_\mathrm{los}^2 (\tau)\rangle-\langle V_\mathrm{los} (\tau)\rangle^2$. 

\section{The simulated discrete radial velocity fields (SDRVF)\label{SDRF}}
The dynamical state of a galactic system is described by its phase
space distribution function (DF). With a functional form for
the density distribution and gravitational potential, it is possible to
determine the DF via the Eddington formula (Binney \& Tremaine \cite{binn})
for spherical systems.

In the case of the adopted Hernquist model,
the Eddington formula is not so useful because its application required
the use of special functions. 
To deal with the DF in a simple heuristic way,
we adopt a factorised form
\be
f({\bf r},{\bf v}) =\rho ({\bf r}) F({\bf r}, {\bf v})
\label{ffatt}
\ee
where $\rho({\bf r})$ is the mass density distribution given by the model
and $F({\bf r, v})$ is the velocity distribution depending on the position,
${\bf r}$, via its moments.

The observed distribution of radial velocity profiles in early-type galaxies
are Gaussian, with maximum deviation of 10\% (Winsall \& Freeman \cite{wins}; Bender
et al. \cite{bend}). In our modeling we will assume a velocity distribution
given by the product of three Gaussian functions, although some anisotropy
is implied by the assumed intrinsic cylindrical rotation.
We write the velocity distribution as
\be
F({\bf r}, {\bf v})=F_R ({\bf r}, v_R ) F_{\varphi} ({\bf r}, v_{\varphi})
F_{z} ({\bf r}, v_{z})
\label{Fvfatt}
\ee
where
\be
F_i ({\bf r}, v_i)=\frac{1}{\sqrt{2\pi}\sigma_i({\bf r})}\exp
\left[ -\frac{(v_i -V_{i}({\bf r}) )^2}{2\sigma_i ^2({\bf r})} 
\right];~i=R,\varphi,z
\label{gaus}
\ee
and $F_R$, $F_{\varphi}$, $F_{z}$ are the normal distributions along
the directions of the principal axes of the velocity ellipsoid. For
each Gaussian, the mean value $V_{i}$ represents the mean motion and
the standard deviation $\sigma_i$ is the diagonal element of the
velocity ellipsoid in the same direction, both evaluated at position
${\bf r}$ $^($\footnote{The DF as given in Eq.~(\ref{ffatt}), with the
assumptions (\ref{Fvfatt}) and (\ref{gaus}) for the velocity
distribution, is not a solution of the collisionless Boltzmann equation
and does not represent the true dynamical state of a stellar system.
However, for our purpose this approximate representation of the DF is
taken as a reasonable description of an instantaneous state for a
spherical system.}$^)$.\\

We consider both non-rotating and rotating systems: once the rotation
velocity structure is assigned, we solve the Jeans equations for the 
given density distribution and obtain the velocity 
dispersion ellipsoid, plus the related projected velocity dispersion profiles.
In rotating systems, the velocity distribution is given by Eq.~(\ref{gaus}),
setting $V_{i}=v_\mathrm{rot}$ as in Eq.~(\ref{rotin})
and $\sigma_i$ as solutions of the Jeans equations Eqs.(\ref{axJ1}),
(\ref{axJ2}); in non-rotating system, $V_{i}=0$ and $\sigma_i$ is given by Eq. (\ref{nrsol})\\

\noindent
We now generate a sample of discrete test particle from the mass density
distribution $\rho({\bf r})$ and the distribution function 
$F({\bf r}, {\bf v})$ 
as follows:
\begin{enumerate}
\item we randomly extract from $\rho({\bf r})$ the spherical coordinates
$(r,\theta, \varphi)$ for each star and check 
that the corresponding projected radius $R$ is greater than $R_\mathrm{lim}$
(see Sect. 2),
\item from the velocity distribution $F({\bf r}, {\bf v})$ we randomly 
extract the three velocity components 
$(v_R, v_{\varphi}, v_{z})$ at the position $ {\bf r}$ of each star.
\end{enumerate}

Once the 3D velocity field is obtained, we project it on the sky plane.
The next step is to simulate a measurement of this two-dimensional 
velocity field: 1) we assume a typical measurement error for low to medium
dispersion multi-object spectroscopy ($\sigma_\mathrm{mea} \sim$ 
70 km/s = 0.08$v_\mathrm{g}$, see footnote 1), with a normal distribution; 2) the 
observed radial velocity measurement is obtained by a random extraction from 
a Gaussian, whose average is the $v_\mathrm{rad}$ value obtained for a single PN 
and standard deviation $\sigma_\mathrm{mea}$.\\
In this way we obtain a simulated discrete radial velocity field, $V_\mathrm{obs}(X,Y)$, from 
which the kinematical information is extracted.

\section{Standard analysis procedure \label{sproc}}
From our simulated systems we want to estimate the kinematical quantities
without using our knowledge of the intrinsic dynamics of our system. 
Therefore we
apply to the simulated data a procedure similar to those adopted
in the analysis of the small samples of discrete radial velocity fields 
of observed PNe
(eg Arnaboldi et al. \cite{arn1}, \cite{arn2}, \cite{arn3}). In this procedure, we fit some 
simple three-parameter functions to our simulated data V$_\mathrm{obs}(X,Y)$.
We do this to see how misleading these fits can be if the
real rotation fields have the form of Eq.~(\ref{rotin}).
We perform a last-squares fit to the V$_\mathrm{obs}(X,Y)$ using:
\begin{enumerate}
\item a {\em bilinear function} (hereafter BF) 
\be
v_\mathrm{rad}=a+bX+cY
\label{bil}
\ee
where $X,\, Y$ are Cartesian coordinates of the PNe (typically in CCD
coordinates) and $a$, $b$, $c$ are parameters to be determined 
by fitting Eq.~(\ref{bil}) to the data.
These parameters are used to determine the systemic velocity, 
direction of the axis of maximum velocity gradient, $Z1$, and modulus 
of the velocity gradient. 
This linear velocity field is equivalent to a solid body rotation of the form 
$v_\mathrm{rad}=v_\mathrm{sys}+\omega r \cos(\phi-\phi_{*})$ where
$v_\mathrm{sys}$ is the systemic velocity, $\omega$ is the angular velocity,
$\phi$ is the angle from the kinematic major axis, and
$\phi_*$, the P.A. of the $Z1$ axis.
\item a {\em flat rotation curve} (hereafter FC) 
\be
v_\mathrm{rad}=v_\mathrm{sys}+v_{*}\cos(\phi-\phi_{*})
\label{flfit}
\ee
where $v_\mathrm{sys}$ and $\phi_*$ are the same quantities as in the BF, and 
$v_*$ is the amplitude of the flat rotation curve.
\end{enumerate}

\noindent
The {\em residual field}, $\Delta v$, is computed as the difference
between V$_\mathrm{obs}$ and the interpolated radial fields from the fits.
The residual fields obviously depend on the adopted fit (BF or FC).
Furthermore we assume point-symmetry to the galaxy center for the
simulated samples. Therefore to every velocity $v$ at position ($X,Y$)
on the sky plane, there is a corresponding velocity $-v$ at position
($-X,-Y$). Hereafter we will refer to a {\em symmetrized velocity
field} when we consider a sample generated on one side of the galaxy by
adding the symmetric points generated by test particles on the other
side.

\section{Simulations \label{sim}}
In Table~\ref{tab1} we summarise the set of parameters used in our
models to generate the SDRVFs. Models with $V_\mathrm{max}$
up to 0.23$v_\mathrm{g}$ are physical, i.e. they have velocity dispersion
ellipsoid components which are positive everywhere, while for
$V_\mathrm{max}>0.23v_\mathrm{g}$ the $\sigma_R= \sigma_\phi$ components become negative for large $z$. However, we
have kept these models because they have physical solutions near the
equatorial plane, where the investigation of the kinematical
observables (rotation velocity and velocity dispersion profiles) is
often focussed. 
\begin{figure*}
\centering
\hspace{-1cm}
\epsfig{figure=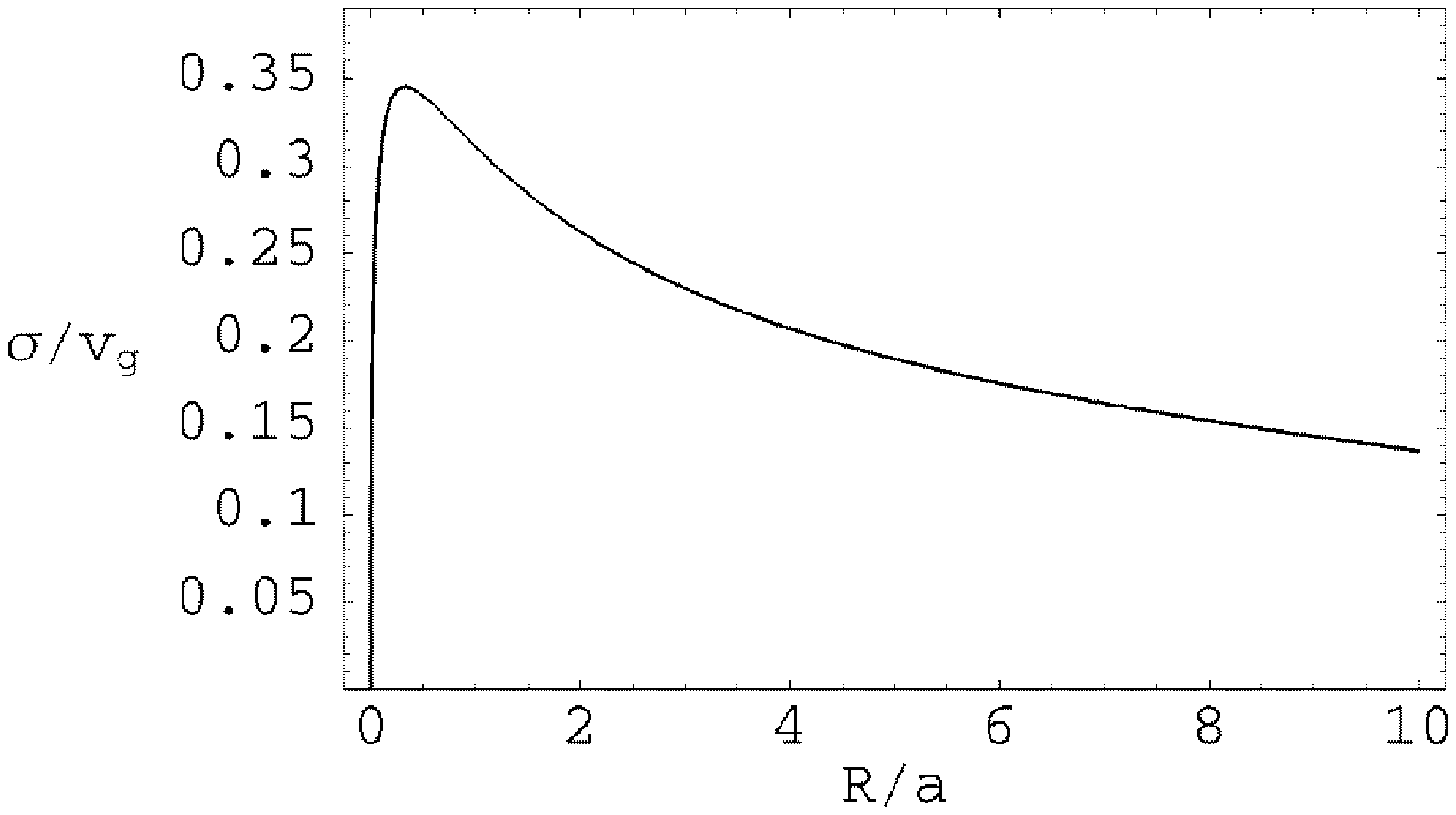,width=7.5cm,height=7cm,angle=-90}
\epsfig{figure=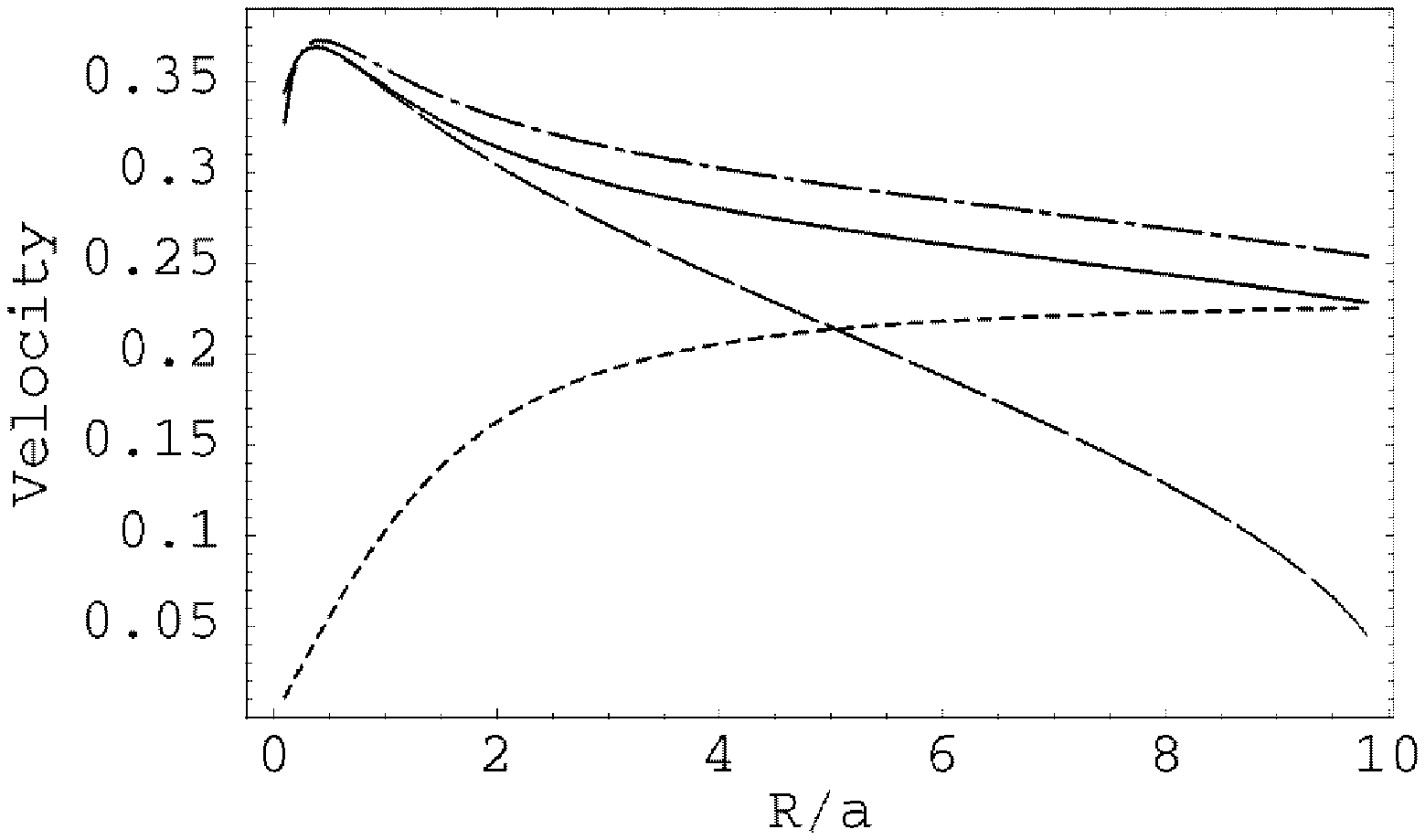,width=7.5cm,height=7cm,angle=-90}
\vspace{-1.2cm}
\caption{\footnotesize Here we show the solutions of the Jeans
equations  (Eqs.~\ref{axJ1} and \ref{axJ2}) as functions of distance 
from the center
(R=$r$, $z$). Left: isotropic non-rotating system without DM. Right:
rotating equilibrium model with DM and $V_\mathrm{max}=0.23v_\mathrm{g}$ (see Table
\ref{tab1}); the $\sigma_R=\sigma_{\varphi}$ component is drawn as
solid line on the equatorial plane and as long-dashed line along the
rotation axis, while the $\sigma_z$ component is the dash-dotted line.
The rotation velocity on the equatorial plane is also plotted, as
short-dashed line.}
\label{selsol}
\end{figure*}

\begin{figure*}
\centering
\epsfig{figure=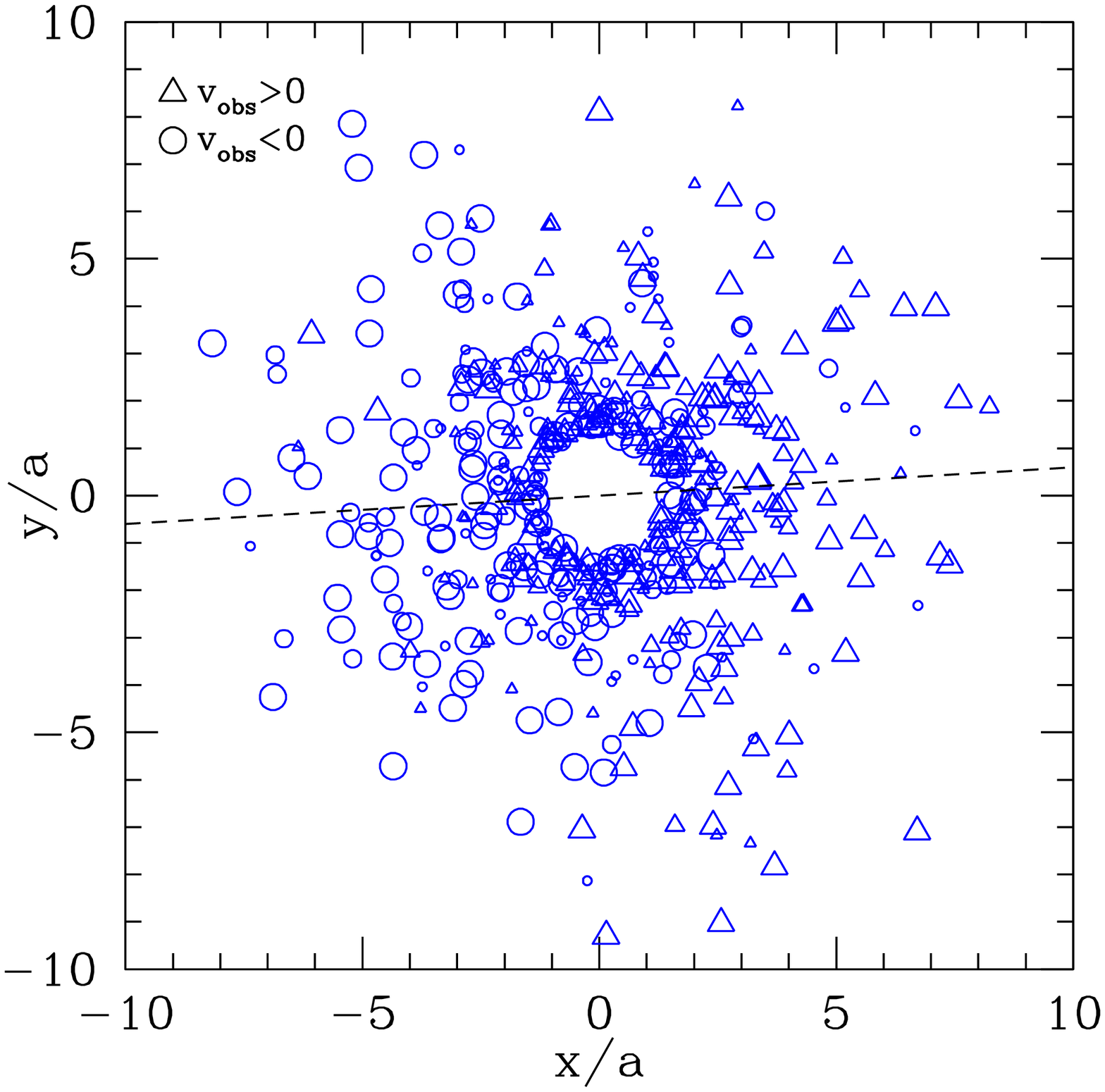,width=6cm,height=6cm}
\vspace{0.5cm}
\epsfig{figure=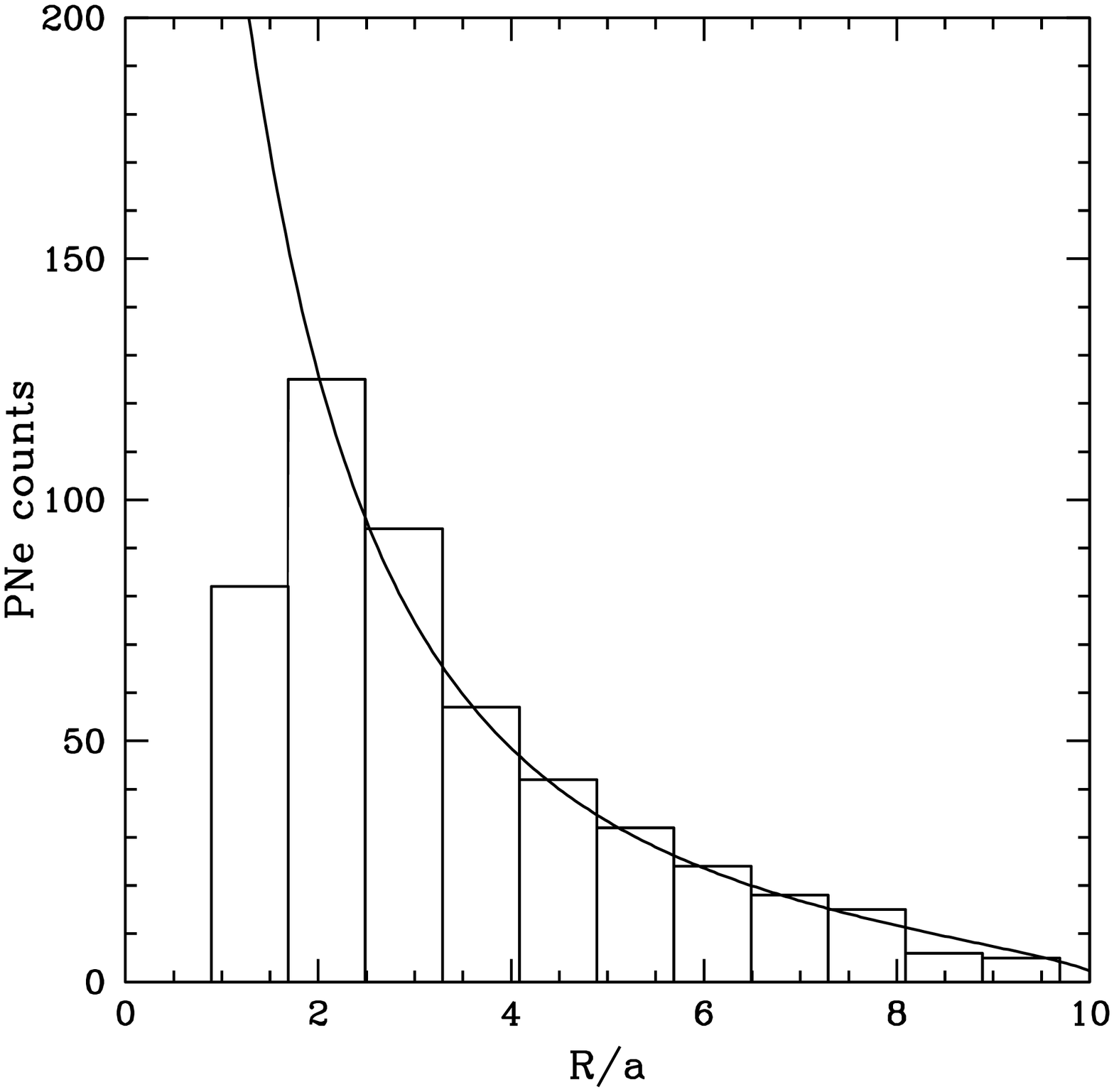,width=6cm,height=6cm}
\caption{\footnotesize Left: a typical SDRVF obtained for a sample
size of 500 PNe and the equilibrium model with DM and cylindrical rotation
(inclination angles: 0:0:0). The dashed line is the line of the maximum 
gradient (computed with the bilinear fit). Right: the PNe number 
density (histogram) is compared with the surface brightness of the simulated 
system (solid line). The incompleteness of the sample is evident in the inner 
regions.}
\label{campo}
\end{figure*}
The solutions of the Jeans equations are shown in Fig.~\ref{selsol}
for the system with $V_\mathrm{max}=0.23v_\mathrm{g}$ and for the non-rotating system. For our models we produce samples with 50, 150 and 500 PNe. For a given sample size, a
fixed mass density model, a particular set of values for the rotation
velocity (as given in Table~1) and inclination of the system, we
produce 100 realisations of the radial velocity fields as described in
Sect.~3. Fig.~\ref{campo} shows a view of the the radial velocity
field and the projected number density distribution obtained for a
sample of 500 PNe for a rotating simulated system with
$V_\mathrm{max}=0.23v_\mathrm{g}$ (see Table~\ref{tab1}) seen edge-on. The comparison
with the luminous matter projected density distribution shows the
expected incompleteness effect in the generated PNe sample towards the
center of the mass distribution. When computing the expected model
quantities from Eq.~(\ref{vlos}), we take this selection effect into
account, using the spatial and projected density profile obtained from
a sample of 50\,000 objects. 

\begin{center}
\begin{table}
\caption{Parameter values for the dynamical models which are used to 
generate the 2-D discrete radial velocity samples. The $a$ and $M_\mathrm{l}$ 
parameters are common to the self-consistent non-rotating and dark matter 
rotating systems.
Here we adopt different values for the rotation vs. velocity dispersion 
so that any bias effects caused by the assumed functional form of the 
rotation field can be investigated.
We refer to Sect. \ref{sim} for a detailed discussion.
In parentheses we show the $v/\sigma$ value at $R=10a$.}
\centering
\begin{tabular}{cc}\hline \hline 
\multicolumn {2}{l}{\normalsize{\footnotesize \bf Dynamical and kinematical
parameters}}\\
\hline \hline \noalign{\smallskip}
$a$ & 1\\
$M_\mathrm{l}$ & 1\\
$d$ & 10$a$\\
$M_\mathrm{d}$ & 7.7$M_\mathrm{l}$ \\
$r_\mathrm{v}$ & 2$a$ \\
\hline
& 0.35 $v_\mathrm{g}$ ($v/\sigma$=1.8) \\
& 0.30 $v_\mathrm{g}$ (1.4) \\
$V_\mathrm{max}$ & 0.23 $v_\mathrm{g}$ (1.0) \\
& 0.20 $v_\mathrm{g}$ (0.8) \\
& 0.15 $v_\mathrm{g}$ (0.6) \\
\hline

\hline
\hline\noalign{\smallskip}
\end{tabular}
\label{tab1}
\end{table}
\end{center}
\subsection{Global kinematical quantities}
To each SDRVF, we apply the analysis procedure described in Sect.
\ref{sproc} and derive the following kinematical quantities:
\begin{enumerate} \item the {\em velocity gradient}, $gradV=\pm
\sqrt{b^2+c^2}$ for the BF, with sign given by the sense of rotation;
\item the {\em maximum rotation velocity} (the $v_*$ parameter) for the
FC; \item the {\em position angle of the maximum gradient},
$\Phi_{Z1}=\arctan(c/b)$ for the BF, or $\Phi_{Z1}=\phi_*$ for the
FC. Hereafter we refer to the axis of the maximum velocity gradient
as $Z1$, with $Z2$ its perpendicular axis, i.e. the {\em apparent
rotation axis}; \item the {\em systemic velocity} $v_\mathrm{sys}=a$ for the
BF and $v_\mathrm{sys}$ parameter from FC. 
\end{enumerate} 
For each set of 100 simulations and a given
sample size, we obtain the distribution of these global quantities, the
related mean value and standard deviation of the sample (SD)\footnote{In this work we will use the abbreviation SD instead of the usual symbol $\sigma$ in order to avoid confusion with the velocity dispersion.}, which are then compared with the expected
values to check for any presence of biases.

\subsection{Projected kinematical quantities}
To investigate the properties of our SDRVFs, we will implement a
binning procedure, because similar strategies were applied to small
observed radial velocity samples (like those for NGC 1399 and NGC 1316) and 
also
in the case of NGC 5128 for which 433 radial velocities were available.
Similar procedures are still adopted when using GCs data (eg Minniti et
al. \cite{minn}; Kissler-Patig et al. \cite{kpal}). In this framework we can also
establish whether there are biases introduced by binning the observed
radial velocity fields.

Once the line of maximum gradient is identified, the symmetrized
velocity field is spatially binned along $Z1$, selecting particles
within a strip along $Z1$. We have adopted strips having different
width $dZ$ in order to check whether there was some dependence of the
kinematical estimates on this parameter: see Table 3. The spatial binning is
done in such a way that the number of particles in each bin is about
10 or more. We also do a radial binning by selecting PNe in radial annuli.
This binning is usually done for radial velocity samples which are
small, assuming spherical symmetry and isotropy\footnote{Here, the analysis on radial bins is performed only for models which are everywhere physical ($V_\mathrm{max}\le 0.23 v_\mathrm{g}$).}. In our analysis of a 
given sample size, the radial and $Z1$ binning are fixed and maintained 
for all the 100 SDRVFs (the same binning is adopted for the $Z2$ axis 
because of spherical symmetry). 

In the standard procedure, the mean values of the radial velocity
sample in spatial bins along $Z1$ provides a measure of the rotation
curve V$_\mathrm{rot}$, while the velocity dispersion is obtained as SD in the
same bins from the residual fields for the BF or FC. This gives an
estimate of the observed dispersion $\sigma_\mathrm{obs}$ from which the
projected velocity dispersion is obtained, using the independently
determined measuring error $\sigma_\mathrm{mea}$, as
$\sigma_\mathrm{V}=\sqrt{\sigma_\mathrm{obs}^2-\sigma_\mathrm{mea}^2}$. This velocity
dispersion obviously depends on the adopted functional form of the
rotation field. For comparison, we also estimated the velocity
dispersion independently of the adopted rotation field, as the SD of the velocities from the mean radial velocity in each bin.
Hereafter we refer to these estimates as NFP (no-fit procedure). We
associate to each value of mean radial velocity V$_\mathrm{rot}$ an error
given by $\delta V_\mathrm{rot}= SD/\sqrt{N}$ and to $\sigma_\mathrm{obs}$ an error $\delta
\sigma_\mathrm{obs}=SD/\sqrt{2N}$. The behavior of $V_\mathrm{obs}$, $\sigma_\mathrm{V}$ and
their relative errors from 100 simulations allow us to estimate the
precision of any observable for a given sample size.

\section{Results \label{res}}
\subsection{Estimates of Kinematical Parameters\label{resGKQ}}
In Table~\ref{tares1} we list the results from our fits for the 
rotating model seen from three different line-of-sights 
(as identified by the Euler angles).
For each set of simulations and fits, Table~\ref{tares1} shows the
expected and estimated values of the kinematical parameters. The
systemic velocity, $V_\mathrm{sys}$, and the P.A. of the maximum velocity
gradient, $\Phi_{Z1}$, are well estimated from the fits of a simple
three-parameter function. They are consistent with the expected model
values at the 95\% confidence level, for the 
potential with and without dark matter, and for both  
functions (BF and FC) that were fit to the velocity
fields. We note the importance of an accurate estimate for
the position angle $\Phi_{Z1}$, because $\Phi_{Z1}$ is
then used in the analysis of the projected kinematics.
Systematic errors on $\Phi_{Z1}$ produce systematic errors on the
projected kinematics. 

\begin{figure*}
\centering
\vspace{-1.5cm}
\epsfig{figure=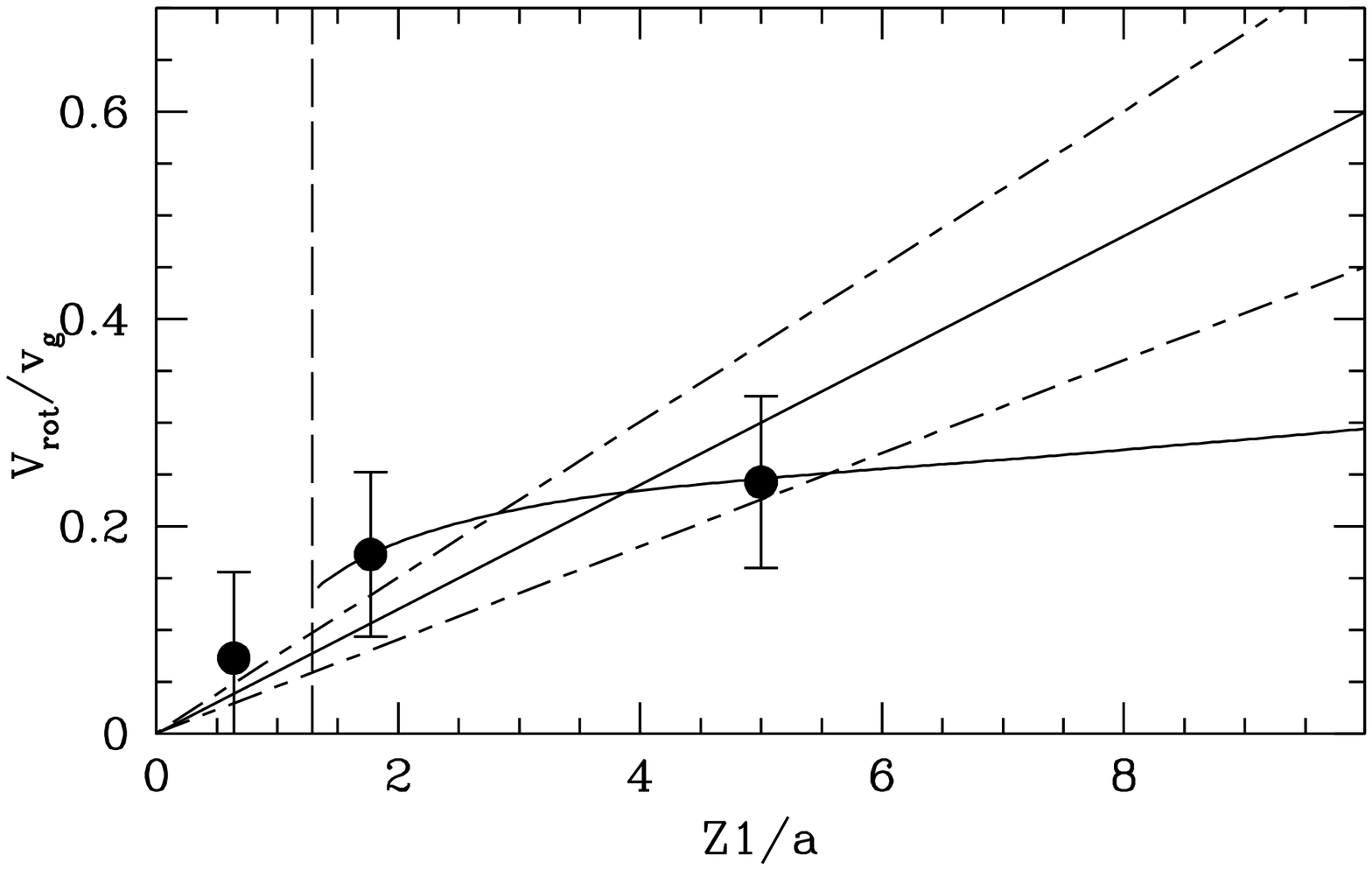,width=6cm,height=6cm}
\epsfig{figure=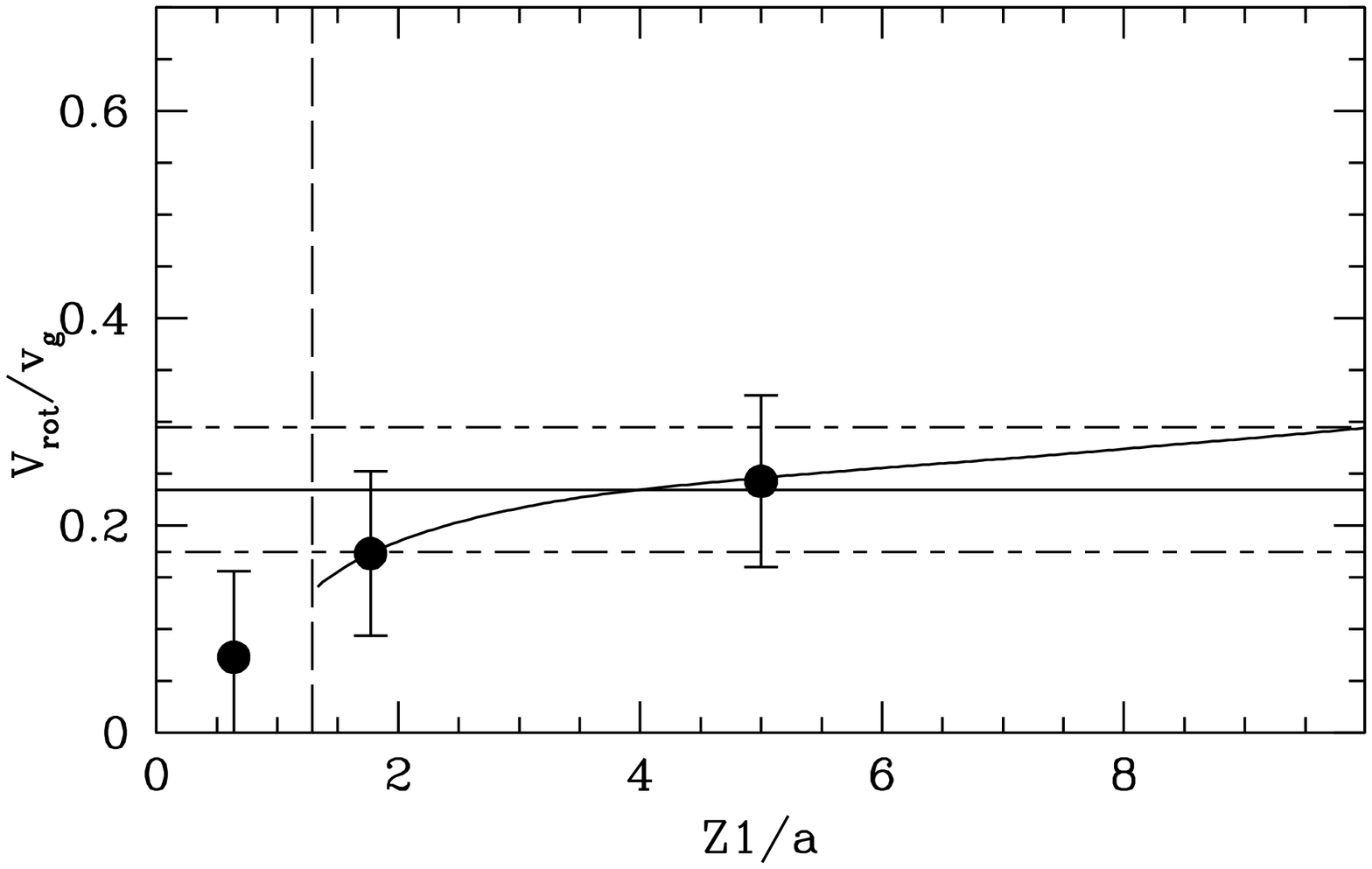,width=6cm,height=6cm}
\caption{\footnotesize Biases in the maximum gradient and maximum rotation 
velocity. The intrinsic rotation curve from the model (curved solid line, $V_\mathrm{max}=0.3v_\mathrm{g}$, see Table \ref{tab1}) is compared with the bilinear fit (left), the flat-curve fit (right) and the mean values in bins (filled points) obtained on 100 simulations. The parameters 
of the fitted functions are the averages from the same 100 simulations:  straight solid lines show the BF (left) and FC (right) fits, and dot-dashed lines are the $\pm$SD lines about the fits. Quoted errors about the points are the SDs of the distribution of estimates in each bin. The dashed vertical line indicates the inner limiting radius, $R_\mathrm{lim}$.}
\label{fitconf}
\end{figure*}

The velocity gradient and maximum rotation velocity estimates are 
systematically incorrect when they are obtained from the wrong simple 
function. In the case of cylindrical rotation, the BF can produce a 
large over-estimate of the velocity gradient (up to about 100\%) while 
the FC (which is closer to being the correct functional form for the
rotation adopted in the model) gives only a small under-estimate. The
reason is clear from examination of Fig.~\ref{fitconf}

{\footnotesize
\begin{table*}
\caption{Results obtained for the equilibrium model
with DM and flat cylindrical rotation: behavior from 100 SDRVFs. 
Square parentheses: the unit in which the quantities are given; round 
parentheses: the expected values. Quoted errors are SDs of
the derived quantities. }
\begin{tabular}{c|ccc|ccc}
\hline\hline
\multicolumn {7}{c}{{\bf Model with DM - Rotation Curve of Equation (9)}}\\ 
\hline\hline
\noalign{\smallskip}
{\bf Proced.} & \multicolumn{3}{c|}{{\bf BF}} & \multicolumn{3}{|c}{{\bf
FC}}\\
\hline
{\bf Inclin.} & \multicolumn{6}{c}{$\phi=0$ $\theta=0$ $\psi=0$} \\
\hline
{\bf $N^o$ PNe} & $\Phi_{Z1}$[rad](0)& gradV$[v_\mathrm{g}/a]$ (0.023) &
$V_\mathrm{sys}[v_\mathrm{g}]$
(0) & $\Phi_{Z1}$[rad](0)& $v_*[v_\mathrm{g}]$ (0.23) & $V_\mathrm{sys}[v_\mathrm{g}]$ (0) \\
\hline
500 & 0.00$\pm$0.08 & 0.046$\pm$0.005 & 0.002$\pm$0.013 & -0.01$\pm$0.09 &
0.182$\pm$0.019 & 0.000$\pm$0.014\\
150 & -0.06$\pm$0.22 & 0.045$\pm$0.010 & 0.01$\pm$0.02 & 0.00$\pm$0.23 &
0.18$\pm$0.04 & 0.00$\pm$0.02 \\
50 & 0.0$\pm$0.3 & 0.048$\pm$0.015 & 0.0$\pm$0.04 & 0.0$\pm$0.3 &
0.18$\pm$0.06 & 0.00$\pm$0.04 \\
\hline
{\bf Inclin.} & \multicolumn{6}{c}{$\phi=-45$ $\theta=-45$ $\psi=-45$} \\
\hline
{\bf $N^o$ PNe} & $\Phi_{Z1}$[rad] (0.62)& gradV$[v_\mathrm{g}/a]$ (0.021) &
$V_\mathrm{sys}[v_\mathrm{g}]$ (0) & $\Phi_{Z1}$[rad](0.62)& $v_*[v_\mathrm{g}/a]$ (0.21) &
$V_\mathrm{sys}[v_\mathrm{g}]$ (0) \\
\hline
500 & 0.61$\pm$0.12 & 0.039$\pm$0.005 & -0.002$\pm$0.013 & 0.62$\pm$0.12
&
0.155$\pm$0.022 & -0.003$\pm$0.014\\
150 & 0.61$\pm$0.23 & 0.040$\pm$0.009 & 0.00$\pm$0.02 & 0.61$\pm$0.23 &
0.16$\pm$0.04 & -0.01$\pm$0.02\\
50 & 0.5$\pm$0.4 & 0.041$\pm$0.013 & 0.00$\pm$0.04 & 0.5$\pm$0.3 &
0.16$\pm$0.05 & 0.00$\pm$0.04\\
\hline
{\bf Inclin.} & \multicolumn{6}{c}{$\phi=90$ $\theta=90$ $\psi=30$} \\
\hline
{\bf $N^o$ PNe} & $\Phi_{Z1}$[rad] (indet.)& gradV$[v_\mathrm{g}/a]$ (0) &
$V_\mathrm{sys}[v_\mathrm{g}]$ (0) & $\Phi_{Z1}$[rad](indet.)& $v_*[v_\mathrm{g}/a]$ (0) &
$V_\mathrm{sys}[v_\mathrm{g}]$
(0) \\
\hline
500 & 0.13$\pm$0.99 &-0.001$\pm$0.006 & -0.002$\pm$0.014 & 0.14$\pm$0.67
&
0.01$\pm$0.01 & -0.002$\pm$0.014\\
150 & 0.01$\pm$0.89 & 0.001$\pm$0.013 & 0.00$\pm$0.03 & 0.06$\pm$0.62 &
0.02$\pm$0.03 & 0.00$\pm$0.03\\
50 & 0.09$\pm$ 0.98 & 0.000$\pm$0.022 & 0.00$\pm$0.05 & 0.18$\pm$0.65 &
0.04$\pm$0.04 & 0.00$\pm$0.05 \\
\hline\hline\noalign{\smallskip}
\end{tabular}
\label{tares1}
\end{table*}
}

\subsection{Estimates of kinematical quantities}
\begin{figure*}
\centering
\vspace{-4.5cm}
\epsfig{figure=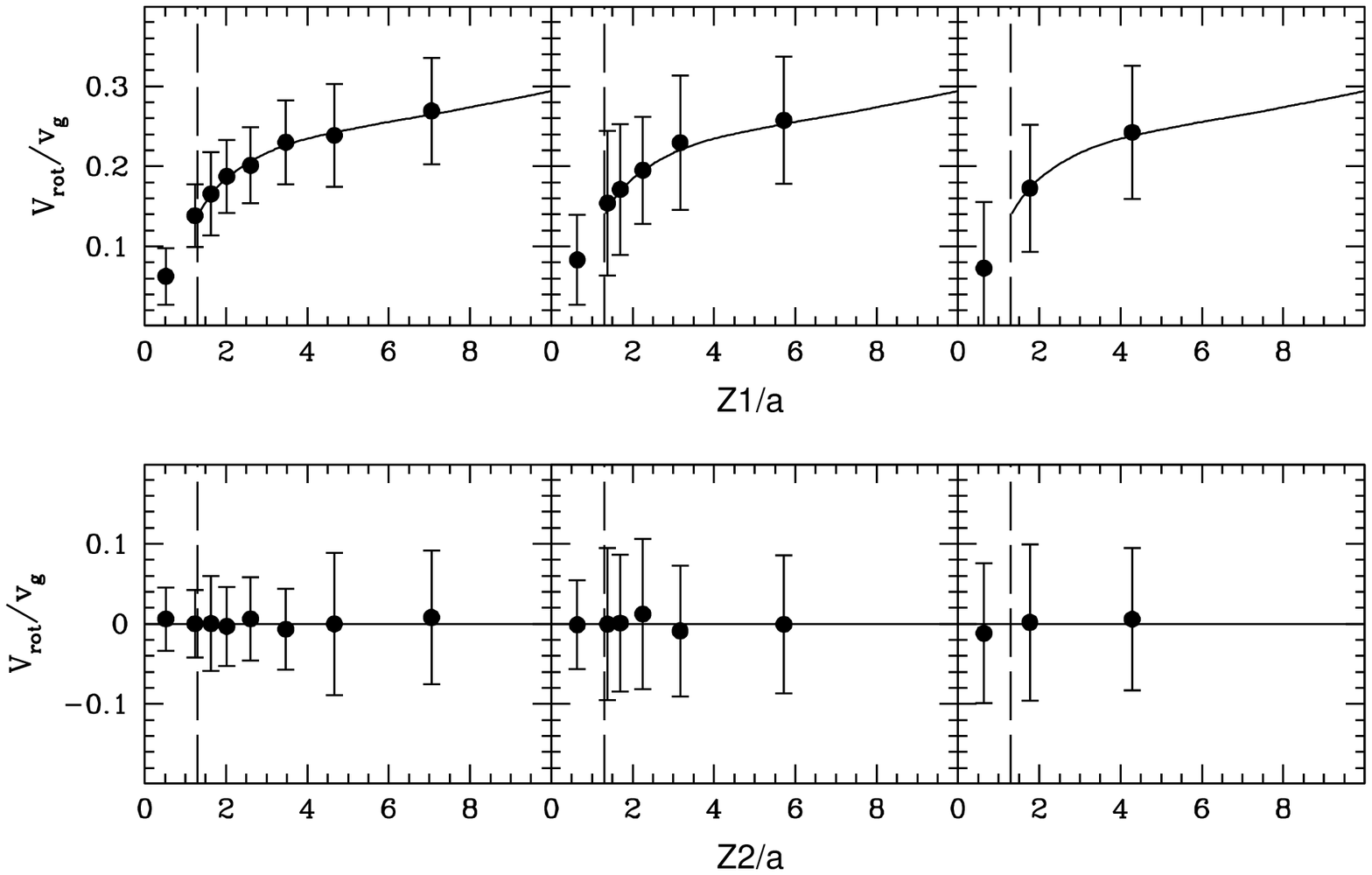,width=14cm,height=14cm}

\centering
\vspace{-9.3cm}
\epsfig{figure=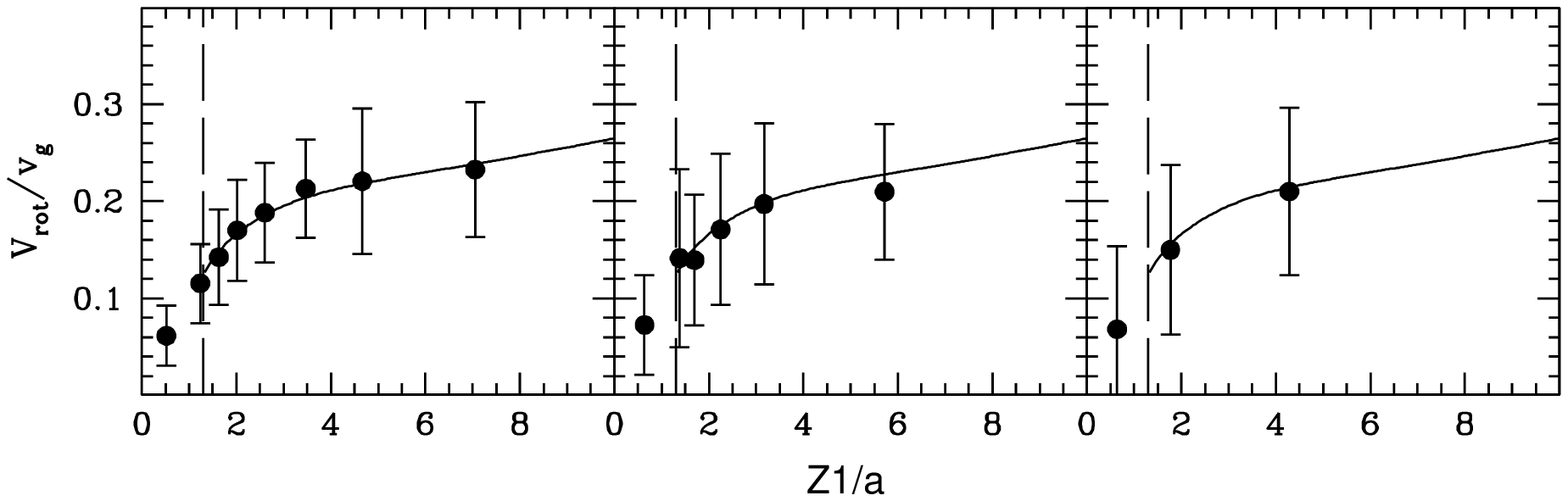,width=14cm,height=14cm}
\vspace{-.5cm}
\caption{\footnotesize Velocity rotation curves for the systems with $V_\mathrm{max}=0.30v_9$. {\bf Top}: Z1 axis ({\em edge-on}); {\bf center}: Z2 axis ({\em edge-on}); {\bf bottom}: Z1 axis (LOS: -45,-45,-45). 
The vertical long dashed line indicates $R_\mathrm{lim}$, full points are the mean values obtained on 100 simulations, solid line is the
expected velocity rotation. {\bf Left}: 500 PNe; {\bf center}: 150 PNe; {\bf
right}: 50 PNe.}
\label{vrot45}
\centering
\vspace{-8.cm}
\epsfig{figure=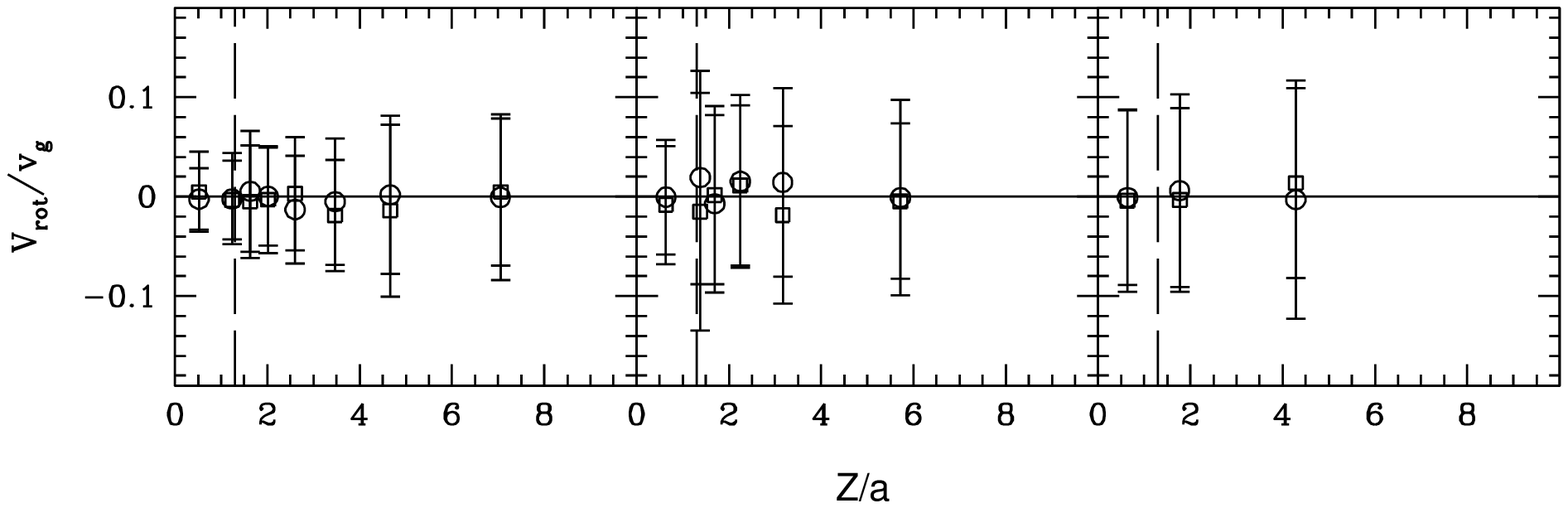,width=14cm,height=14cm}
\vspace{-0.5cm}
\caption{\footnotesize Velocity rotation curves along Z1 and Z2 for the systems with $V_\mathrm{max}=0.30v_9$, {\em face-on} case. Empty
circles are the mean values obtained on 100 simulations along Z1, empty squares are the mean values along Z2, solid line is the
expected velocity rotation. {\bf Left}: 500 PNe; {\bf center}: 150 PNe; {\bf
right}: 50 PNe.}
\label{vrot90}
\end{figure*}

\noindent
{\it Rotation velocity curves} 
The mean velocity rotation curves evaluated from 100 SDRVFs for each given
sample size follow the theoretical behaviour predicted by the models
for every parameter choice (Fig.~\ref{vrot45}).
Rotation along $Z2$ is everywhere consistent with the null rotation
(Fig.~\ref{vrot45}).\\
In Figs~\ref{vrot45} and~\ref{vrot90}
we show the results obtained considering the inclined system:
the rotation velocity is decreased by the expected $\sin i$ factor,
where $i$ is the angle between the line-of-sight and rotation axis
given by a particular set of Euler angles. For {\em face-on} model,
no-rotation is found, as expected (Fig. \ref{vrot90}).\\
~\\
{\it Velocity dispersion profiles along the Z1 and Z2 axes} - 
The inferred velocity dispersion profiles are of particular interest
because they allow the evaluation of the mass distribution via
inversion of the Jeans equations. When the velocity dispersion
profiles are computed from the residual velocity fields, they can be 
affected by biases resulting from the simple functional forms used to 
fit the discrete radial velocity sample.

When the equilibrium model with cylindrical rotation is compared with
the results for larger samples (eg 150 and 500), the velocity dispersion 
profiles obtained from the BF residuals along the Z1 axis show a small
but visible over-estimate of $\sigma_\mathrm{V}$ in the outer bins relative
to the expected values from the model. Better agreement is found
using the FC (see Fig.~\ref{self_cyl20}). This happens because, along
the maximum gradient axis, the flat rotation curve gives a better
description for the rotational structure of the equilibrium model under
study, whose rotation becomes flat at large distances from the center
along the $Z1$-direction. 

\begin{figure}
\vspace{-2.cm}
\centering
\epsfig{figure=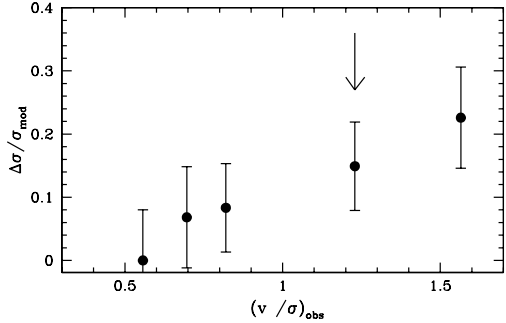,width=7cm,height=7cm}
\caption{Biases from BF in dark matter systems: points represent the relative deviation of the velocity dispersion determined in the last bin along Z1 with respect the ``true'' value expected from model. Error bars take in to account errors on the estimates on 100 simulations. The behavior shown by the residuals indicates that higher 
rotation in the model causes the bilinear fit to introduce a larger bias 
in the estimated velocity dispersion profile. The vertical arrow shows 
the bias in the velocity dispersion estimate, for the model with 
$V_\mathrm{max}=0.30v_\mathrm{g}$ (see also Figure~\ref{self_cyl20}).}
\label{bias}
\end{figure}

We find a clear and expected correlation of this bias with the 
$v/\sigma$ ratio. In the last bin along $Z1$, we compute the difference 
$\Delta \sigma=\sigma_\mathrm{V}-\sigma_\mathrm{mod}$, between the dispersion 
values estimated from BF ($\sigma_\mathrm{V}$), and the expected dispersion from the models ($\sigma_\mathrm{mod}$), for different values of 
$(v/\sigma)_\mathrm{obs}=V_\mathrm{rot}/\sigma_\mathrm{V}$ values in the last bin (obtained by varying the $V_*$ parameter in the rotating models).
The results for 500 PNe are shown in Fig.~\ref{bias}. The bias in the 
velocity dispersion in the last bins is an increasing function of 
the $(v/\sigma)_\mathrm{obs}$ ratio: larger rotation causes the bilinear fit 
to introduce a larger overestimate of the velocity dispersion profile.
 
As mentioned earlier, for small radial velocity samples this kind of
bias is associated not only with our simple parametric fits but also
with more sophisticated non-parametric analyses. When there are fewer
data points, the non-parametric algorithms with their inherent smoothing
effectively fit a plane through the data, so the velocity dispersion
profiles derived from the residual fields do suffer from similar biases
as the one we are investigating in the case of a simple bilinear fit.

\begin{figure*}
\vspace{-0.5cm}
\centering
\epsfig{figure=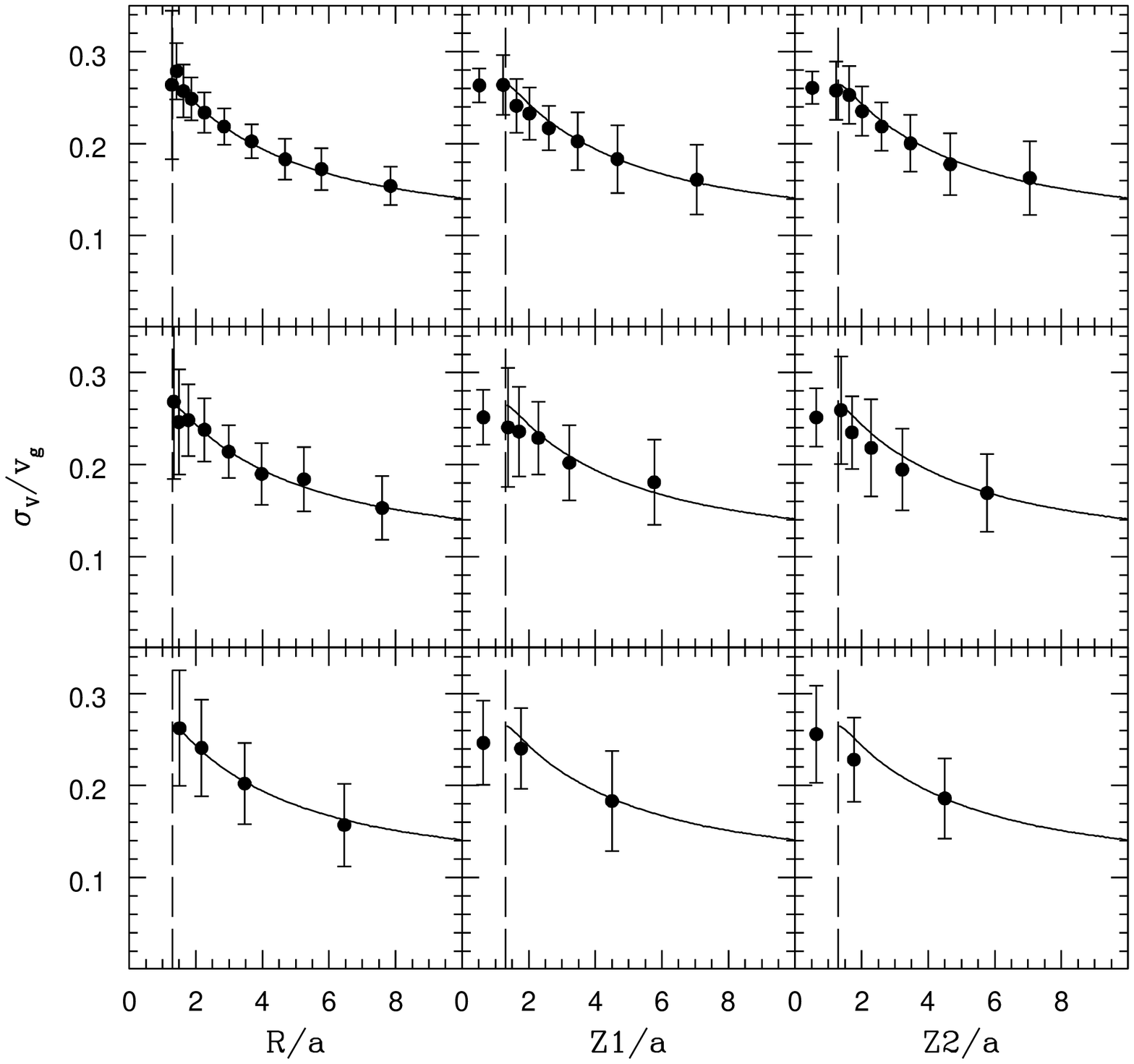,width=12cm,height=12cm}
\vspace{-0.5cm}
\caption{ \footnotesize Velocity dispersion profiles for non-rotating systems: model without DM. Full points are the mean values obtained on 100 simulations, solid line is the expected velocity dispersion profile. {\bf Top}: 500 PNe;{\bf middle}: 150 PNe; {\bf bottom}: 50 PNe. {\bf Left}: in radial bins; {\bf center}: along Z1 axis; {\bf right}: along Z2 axis.}
\label{self_nor}
\vspace{-4cm}
\centering
\epsfig{figure=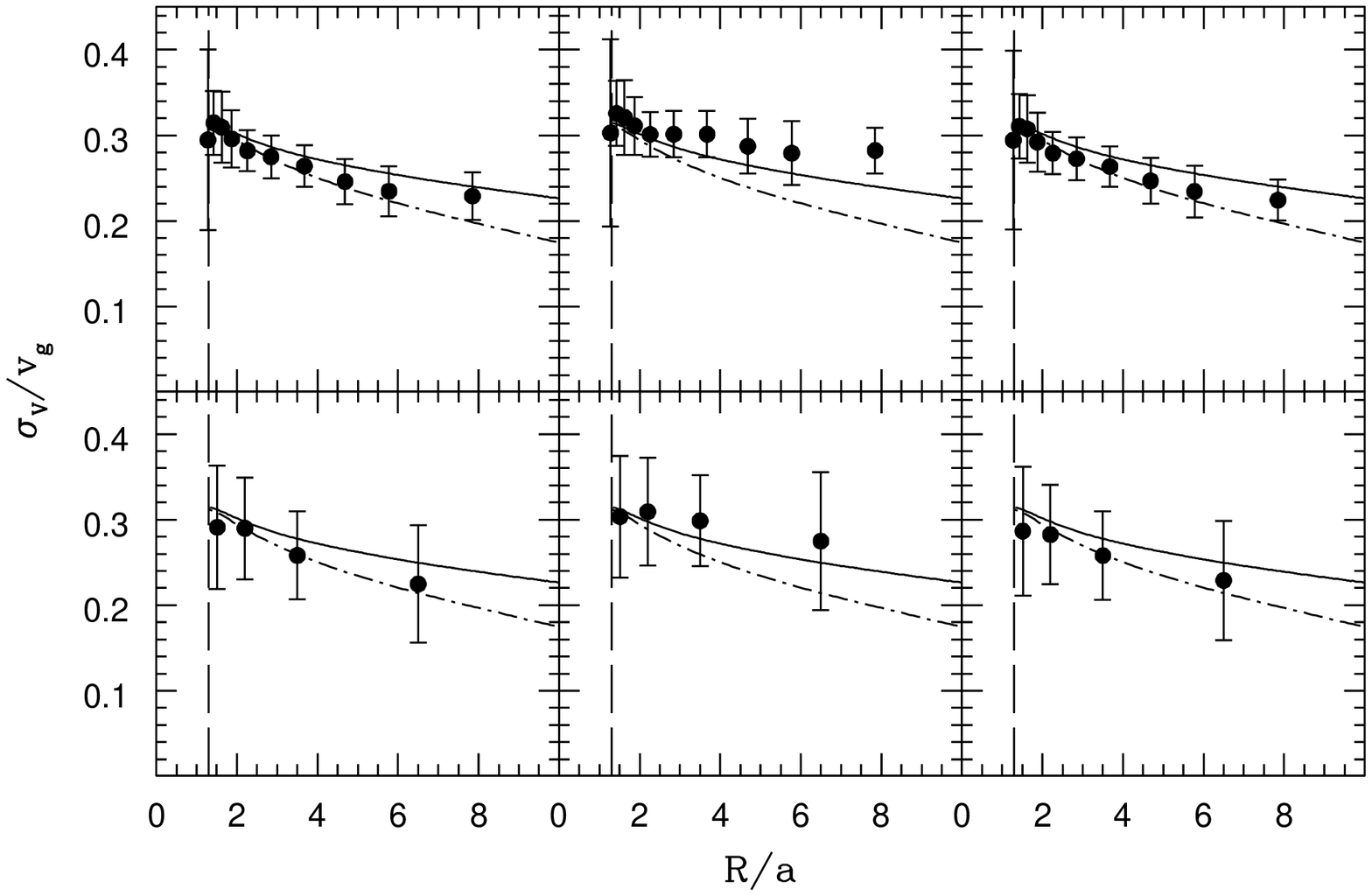,width=12cm,height=12cm}
\vspace{-0.5cm}
\caption{ \footnotesize Velocity dispersion profiles in radial bins:
rotating model with DM and $V_\mathrm{max}=0.23v_\mathrm{g}$, edge-on case.
Full points are the mean values obtained on 100 simulations, solid line is
the
expected velocity dispersion profile along the Z1-axis, dot-dashed line is
the
expected velocity dispersion profile along the Z2-axis. The estimates by the fitting procedures (BF and FC) show the azimuthal dependence while the NFP causes an overestimation of the velocity dispersion values. {\bf Top}: 500 PNe; {\bf
middle}: 150 PNe, {\bf bottom}: 50 PNe. {\bf Left}: bilinear fit procedure;
{\bf
center}: no-fit procedure; {\bf right}: flat-curve procedure.}
\label{self_cylR0}
\end{figure*}

\begin{figure*}
\vspace{-4.5cm}
\centering
\epsfig{figure=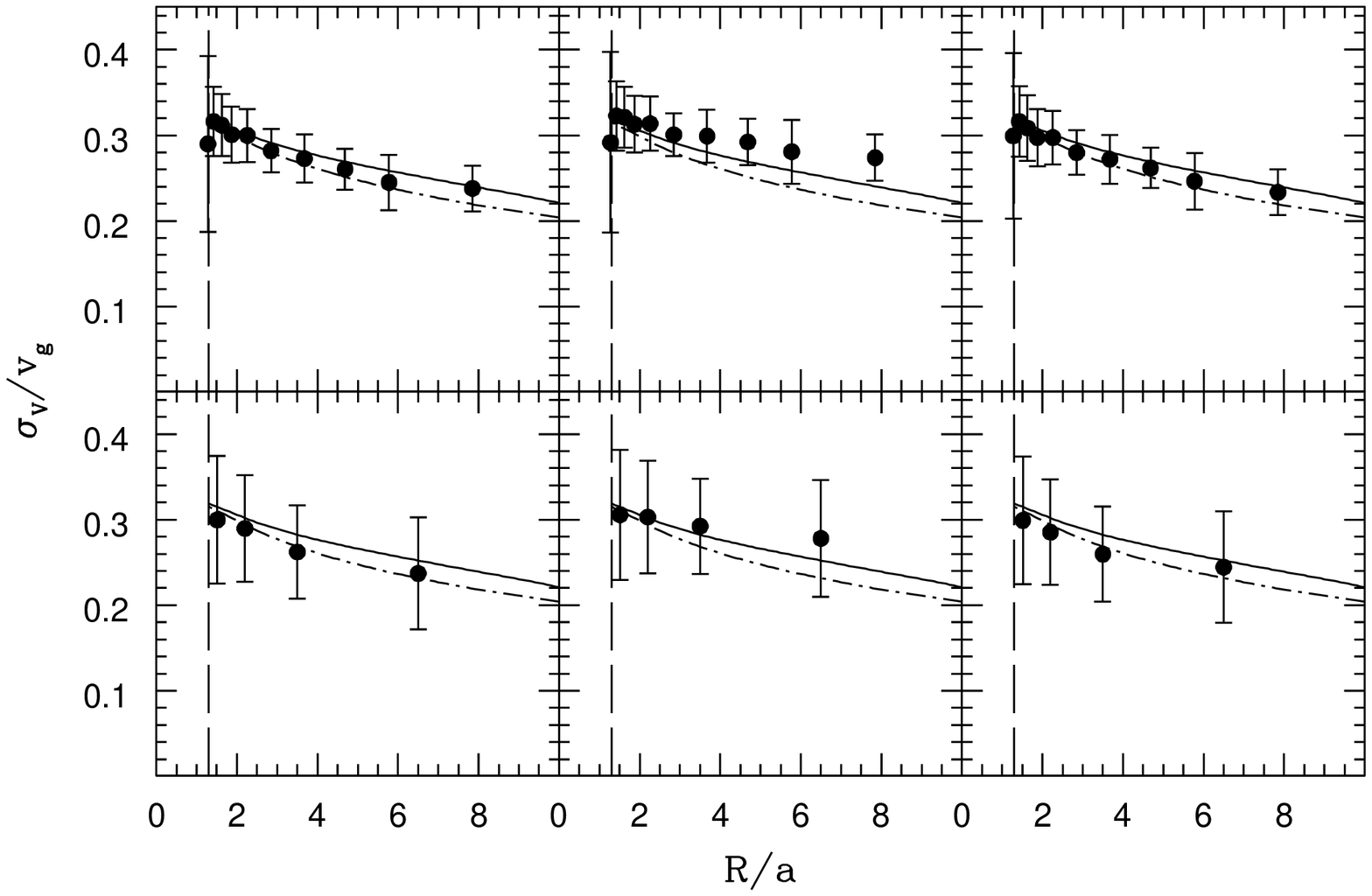,width=13cm,height=13cm}
\vspace{-0.5cm}
\caption{ \footnotesize Velocity dispersion profiles in radial bins:
rotating model with DM and $V_\mathrm{max}=0.23v_\mathrm{g}$. LOS:
-45:-45:-45. Symbols have the same meaning as in
Figure~\ref{self_cylR0}. {\bf Top}: 500
PNe;
{\bf bottom}: 50 PNe. {\bf Left}: bilinear fit procedure;
{\bf
center}: no-fit procedure; {\bf right}: flat-curve procedure.}
\label{self_cylR45}
\vspace{-0.7cm}
\centering
\epsfig{figure=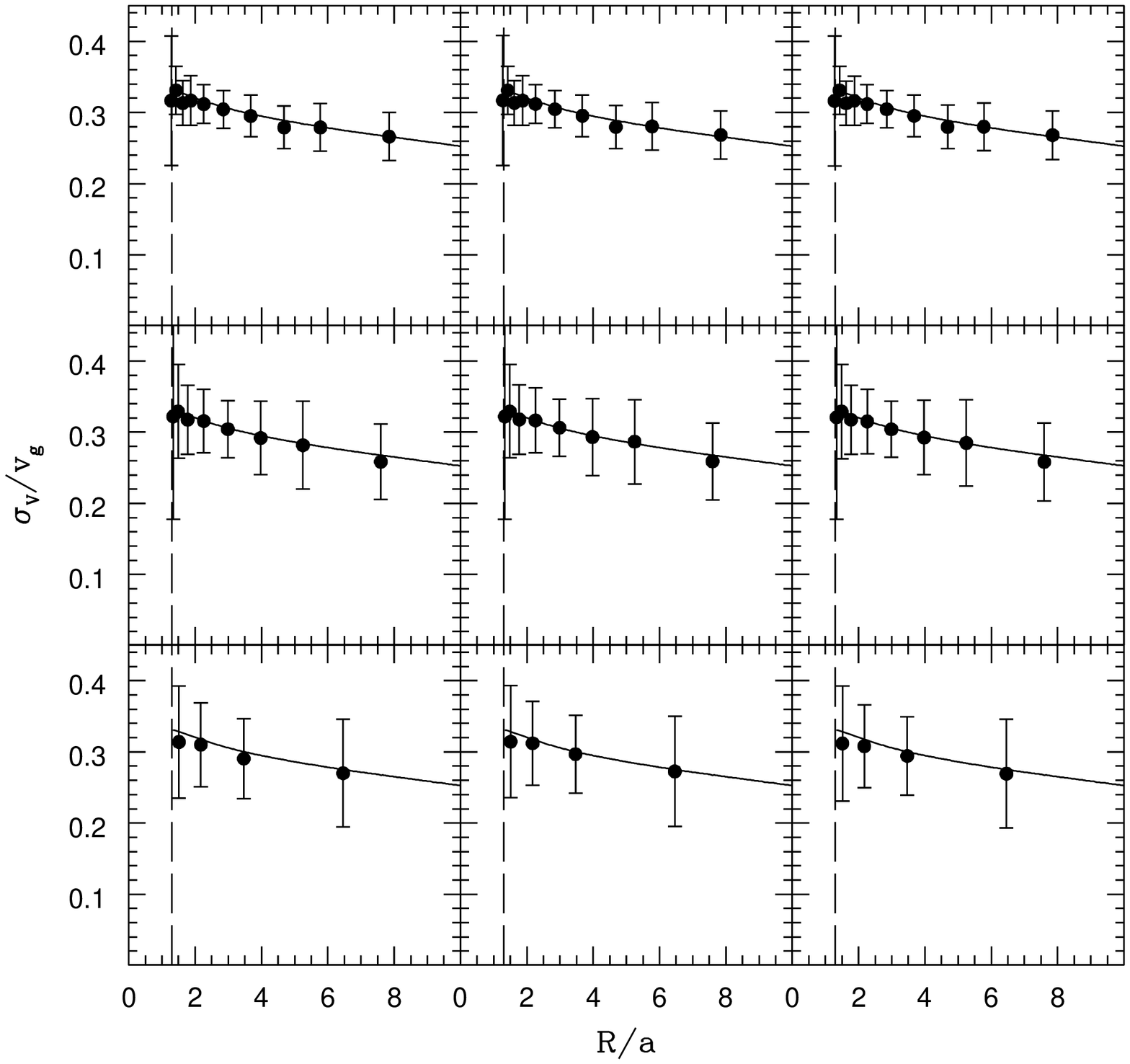,width=13cm,height=13cm}
\vspace{-0.5cm}
\caption{ \footnotesize Velocity dispersion profiles in radial bins:
rotating model with DM and $V_\mathrm{max}=0.23v_\mathrm{g}$, {\em face-on case}.
Here the expected velocity dispersion profile (solid line) has no azimuthal dependence and a perfect agreement is found with the estimates (filled
squares). {\bf Top}: 500
PNe;{\bf center}: 150
PNe;
{\bf bottom}: 50 PNe. {\bf Left}: bilinear fit procedure;
{\bf
center}: no-fit procedure; {\bf right}: flat-curve procedure.}
\label{self_cylR90}
\end{figure*}

In general, for all sample sizes, the velocity dispersion estimates
along the Z1 axis obtained using the NFP are in better agreement with
the expected ones. The estimates along Z2 axis are not affected by
this bias: in Fig.~\ref{self_cyl20} the estimates from BF and FC are
in agreement with the model, while NFP systematically overestimates the
velocity dispersion values in the bins. This effect depends on the 
strip dimension adopted to select particles along the rotational axis: 
the expected profile is computed along the Z2-axis, while the estimates 
are obtained on bins and their widths include regions where  
the velocity dispersion is larger (see Fig.~\ref{cont}). 
On average, they trace a larger dispersion and this causes an overestimate. 
On the other hand, the different fit procedures subtract a gradient in 
these bins, and produce a lower dispersion.
In Fig.~\ref{self_cyl20}, the estimates with the smallest dZ2, 
adopted for the different samples, show the magnitude of this effect. 
Smaller dZ2 reduce the overestimate but the errors are larger. 
All these features are found for the inclined systems too, 
as shown in Fig.~\ref{self_cyl245}. Here the profiles follow those of 
the {\em edge-on} case, but the amount of bias due to the fitting procedure 
is smaller because the $(v/\sigma)_\mathrm{obs}=V_\mathrm{rot}/\sigma_\mathrm{V}$ ratio is 
lower\footnote{The net effect of the inclination is to decrease 
the $(v/\sigma)_\mathrm{obs}=V_\mathrm{rot}/\sigma_\mathrm{V}$ ratio in the simulated systems.}.
In the {\em face-on} case, no biases are found as in case of non-rotating self consistent model (Fig. \ref{self_nor}).\\
~\\
{\it Velocity dispersion profiles from radial bins} -
When we derive the velocity dispersion profiles from the residual
velocity field (either from BF or FC) in {\it radial} annuli, they show
effects from the azimuthal dependence of the intrinsic velocity
dispersion ellipsoid. Those profiles derived either from BF or FC
have a behavior which is intermediate between those expected from the
model along $Z1$ and $Z2$ (Figs. \ref{self_cylR0} and \ref{self_cylR45}) 
for the {\em edge-on} and the inclined cases. 
In the {\em face-on} case (Fig. \ref{self_cylR90}), no
azimuthal dependence is expected or seen (see Fig.~\ref{cont}).
On the other hand, the velocity dispersion profiles obtained
with radial binning and NFP show a strong overestimate of the velocity 
dispersion values due to the cylindrical structure of the intrinsic 
velocity field.

\section{Sampling errors}
The sampling errors are related to the size of a given sample,
and affect the accuracy of the kinematical quantities we are trying 
to estimate. They are important because they often dominate the error 
budget of the estimated mass and angular momentum estimates in halos of 
galaxies. In this work, for given sample sizes, we simulated 100 
realisations of the radial velocity field, from which we derived the 
distribution of the kinematical measurements. We then compare the 
uncertainty estimated (internally) from single measurements of a 
kinematical observable with the uncertainty from the distribution 
of that observable from 100 simulations. This allows us to check
i) whether the statistical errors are evaluated in a realistic way for
single measurements
and ii) study the behavior of the sampling errors and the precisions based
on their actual distributions.
{\footnotesize \begin{table}[h]
\caption[]{Average relative precision considering all bins and models: 
RB are radial bins, $dZ_\mathrm{min}$ and $dZ_\mathrm{max}$
are the minimum and maximum width of the strips adopted along the Z-axes.  
In parentheses we indicate the mean number of PNe selected with our choice
of the bin dimensions.}
\begin{tabular}{llll}
\hline\hline\multicolumn {4}{l}{\footnotesize{\bf Relative errors on
velocity
dispersion}}\\
\hline \hline\noalign{\smallskip}
N$^o$ PNe & RB $(\overline{N})$ & $dZ_\mathrm{min}~(\overline{N})$ & $dZ_\mathrm{max}~(\overline{N})$\\
\hline
500 & 0.106$\pm$0.001 (55) & 0.211$\pm$0.015 (23) & 0.164$\pm$0.006 (34)\\
150 & 0.188$\pm$0.005 (20) & 0.310$\pm$0.015 (9) & 0.233$\pm$0.007 (14)\\
50 & 0.238$\pm$0.016 (15) & 0.38$\pm$0.03 (8) & 0.251$\pm$0.015 (12)\\
\hline\hline\noalign{\smallskip}
\end{tabular}
\label{prec}
\end{table}}
{\footnotesize
\begin{table}[h]
\hspace{-1cm}
\caption[]{Average relative precision in the last bin: RB are radial bins,
$dZ_\mathrm{min}$ and $dZ_\mathrm{max}$ are the minimum and maximum width of the strips 
adopted along the Z-axes. In parentheses we indicate the mean number of PNe 
selected with our choice of the bin dimensions.}
\begin{tabular}{llll}
\hline \hline \multicolumn {4}{l}{\footnotesize{\bf Relative errors on velocity dispersion in last bin}}\\
\hline \hline \noalign{\smallskip}
N$^o$ PNe & RB$~(\overline{N})$ & $dZ_\mathrm{min}~(\overline{N})$ &
$dZ_\mathrm{max}~(\overline{N})$\\
\hline
500 & 0.12$\pm$0.02 (48) & 0.32$\pm$0.03 (8.5)& 0.24$\pm$0.03 (12.5)\\
150 & 0.21$\pm$0.02 (17) & 0.37$\pm$0.04 (8)& 0.25$\pm$0.03 (13)\\
50 & 0.28$\pm$0.02 (11) & 0.47$\pm$0.08 (7) & 0.29$\pm$0.03 (11)\\
\hline\hline\noalign{\smallskip}
\end{tabular}
\label{prec2}
\end{table}}

\subsection{Errors on the velocity dispersion}
For all models and velocity dispersion profiles, we compute the average 
of the relative errors on 100 simulations in all bins (for which the PNe 
number density is complete) for a given sample size. {\it We consider this
to 
be indicative of the precision for the velocity dispersion}. We do this
for the spatial binning along $Z1$, $Z2$, and for radial bins.
The results are shown in Table ~\ref{prec}.
\begin{figure*}
\centering
\epsfig{figure=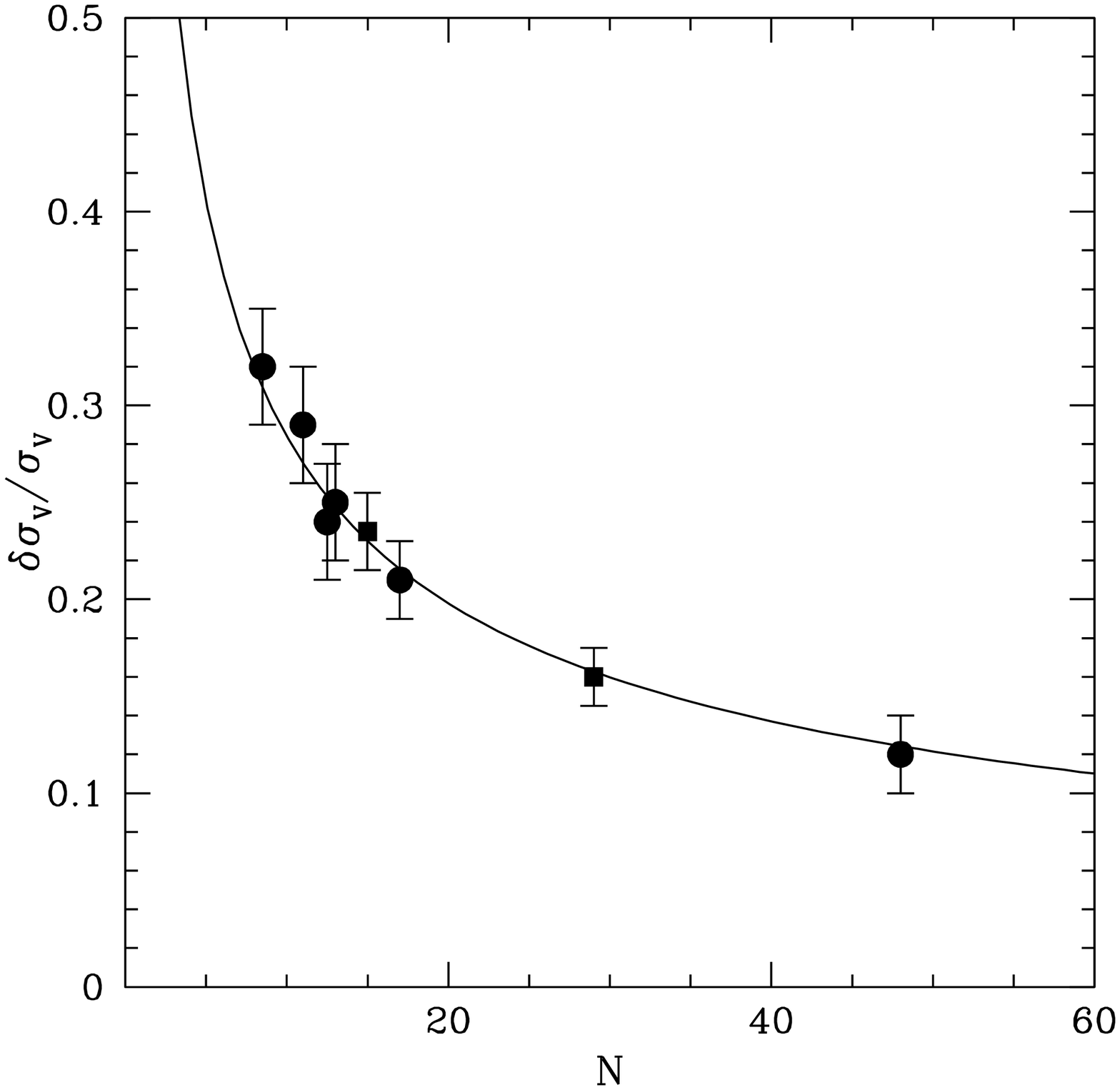,width=6.2cm,height=6.2cm}
\epsfig{figure=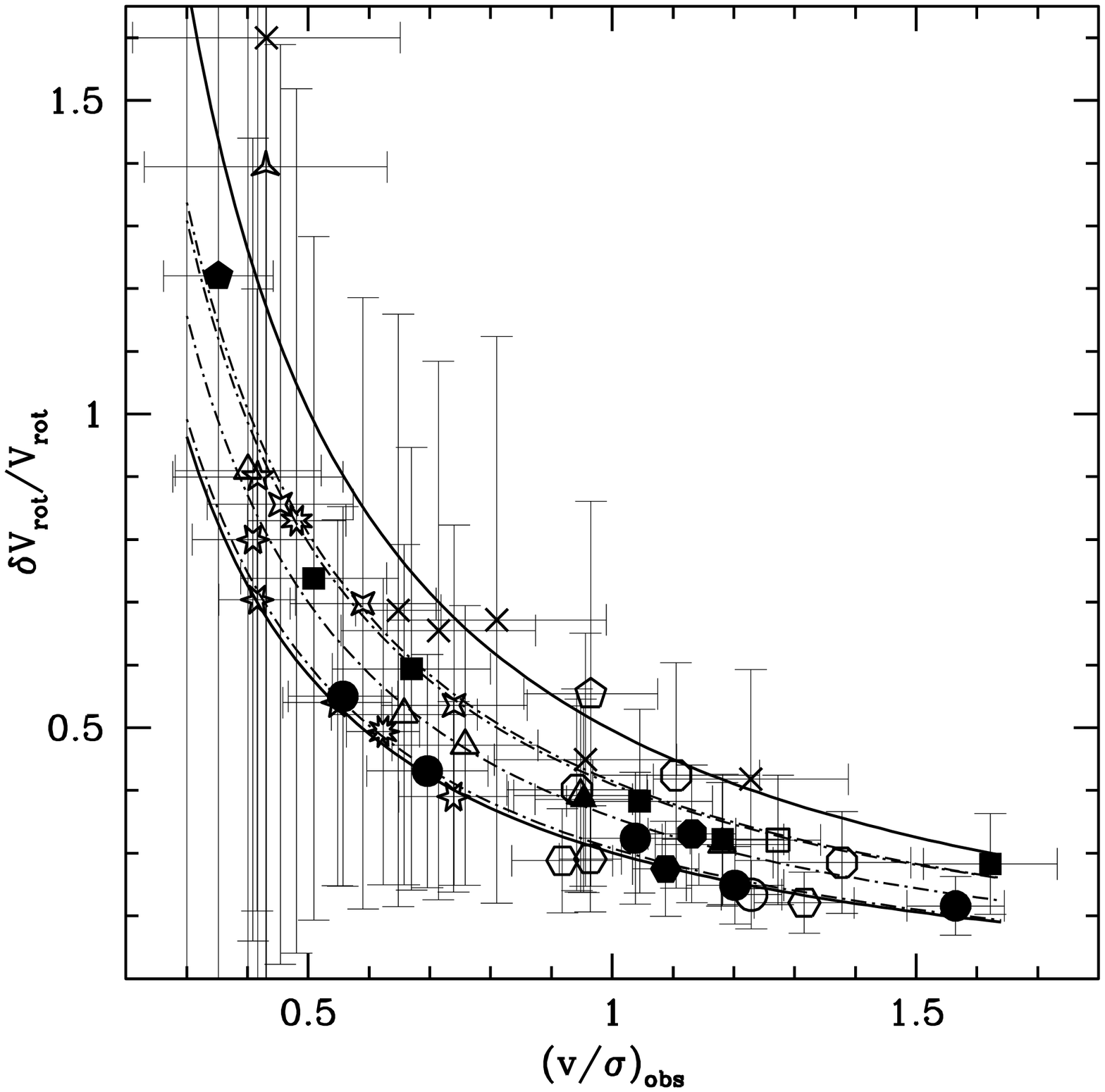,width=6cm,height=6cm}
\caption{\footnotesize Left: behavior of the relative errors on the velocity dispersion with respect the mean number density in the last bin (full 
points) and in two more bins (see footnote 9) on all the simulation done 
for all the models. Solid line is a fit to the data points 
using the function {\footnotesize $A/\sqrt{\overline{N}}+B$}. Right: correlations between relative errors for the rotation velocity and $(v/\sigma)_\mathrm{obs}$ in the last bin. Here different symbols are adopted for different models, sample sizes and bin dimensions. They fill in the region identified by the fitted function for 
the maximum mean sample, $\overline{N}$=13, and the minimum one, 
$\overline{N}$=7, per bin used in our simulations. Dash-dotted lines 
are for intermediate number samples.}
\label{reldi}
\end{figure*}

The average relative errors depend only on sample size and are 
quite independent on the fitting function, inclination, and 
the mass and rotation models\footnote{The variation of the precisions 
between models is less than 10\% for a given sample size.}. 
If we consider the estimates along the Z-axes, for a sample size of 500 
test particles the precision on velocity dispersion is 16\%, for 150 test 
particles it is 23\%, and for a sample of 50 test particles it is 25\%. 
Higher precisions (10\% for 500 PNe, 19\% for 150 PNe, 24\% for 50 PNe) 
are found if we consider the estimates in radial bins 
(where the whole sample size is used). The mean number of particles 
related to these estimates are shown in Table~\ref{prec}.

The results for the last bins are shown in Table~\ref{prec2}. The
relative errors are plotted in Fig.~\ref{reldi} against
$\overline{N}$\footnote{Here are plotted the results from Table~\ref{prec2} 
and additional two values obtained for the 150 PNe in the $4^{th}$ bin  
and for 500 PNe in the $6^{th}$bin.  These values for N=14 
and N=29 respectively are shown with full squares in 
Fig.~\ref{reldi}.}, showing a pseudo-Gaussian behavior. 
The distribution is reproduced by the function $A/\sqrt{\overline{N}}+B$, 
with best fit parameters, A=$0.94\pm0.06$ and B=$-0.01\pm0.02$. 
If $\delta \sigma_\mathrm{V}/\sigma_\mathrm{V}=1/\sqrt{2\overline{N}}$, 
the B and A values should be zero and 
$1/\sqrt{2}=0.71$ respectively. Our estimate for B is consistent
with zero. We find A=0.94$>$0.71 because by definition $\delta
\sigma_\mathrm{p}/\sigma_\mathrm{p}=1/\sqrt{2\overline{N}}$ and 
$\sigma_\mathrm{V}^2=\sigma_\mathrm{p}^2-\sigma_\mathrm{mea}^2$ so
\begin{equation}
\frac{\delta \sigma_\mathrm{V}}{\sigma_\mathrm{V}}=
\frac{1}{1-\sigma_\mathrm{mea}^2/\sigma_\mathrm{p}^2}\frac{\delta \sigma_\mathrm{p}}{\sigma_\mathrm{p}}=
C_{\sigma}\frac{\delta \sigma_\mathrm{p}}{\sigma_\mathrm{p}}>\frac{\delta \sigma_\mathrm{p}}{\sigma_\mathrm{p}}=
\frac{1}{\sqrt{2\overline{N}}}.
\label{err}
\end{equation}
For a typical $\sigma_\mathrm{mea}/\sigma_\mathrm{p}$=0.4 in last bins, 
$C_{\sigma}$=1.2 and the expected value for the A parameter is
A=$C_{\sigma}/\sqrt{2}$=1.2$\times$0.71=0.85. 
Furthermore $\sigma_\mathrm{V}=(\sigma_\mathrm{p}^2-\sigma_\mathrm{mea}^2)^{1/2}$ 
is not a Gaussian variable, as we do not have a linear 
relation between $\sigma_\mathrm{V}$ and $\sigma_\mathrm{p}$, so this causes
an overestimate when $\sigma_\mathrm{V}$ is computed as a Gaussian quantity.\\
Finally, the computed value for A accounts for an average 
behavior of the velocity dispersion relative errors on a 
wide range of models and it can be considered as an upper limit for
the relative errors, once given the number sample in bins. \\
We have then compared the precisions on a single measurement,
obtained as in Eq.~(\ref{err}), with those expected from the 
related distribution on 100 simulations and found
that they are consistent within the errors. 
\begin{figure*}
\centering
\hspace{-2cm}
\epsfig{figure=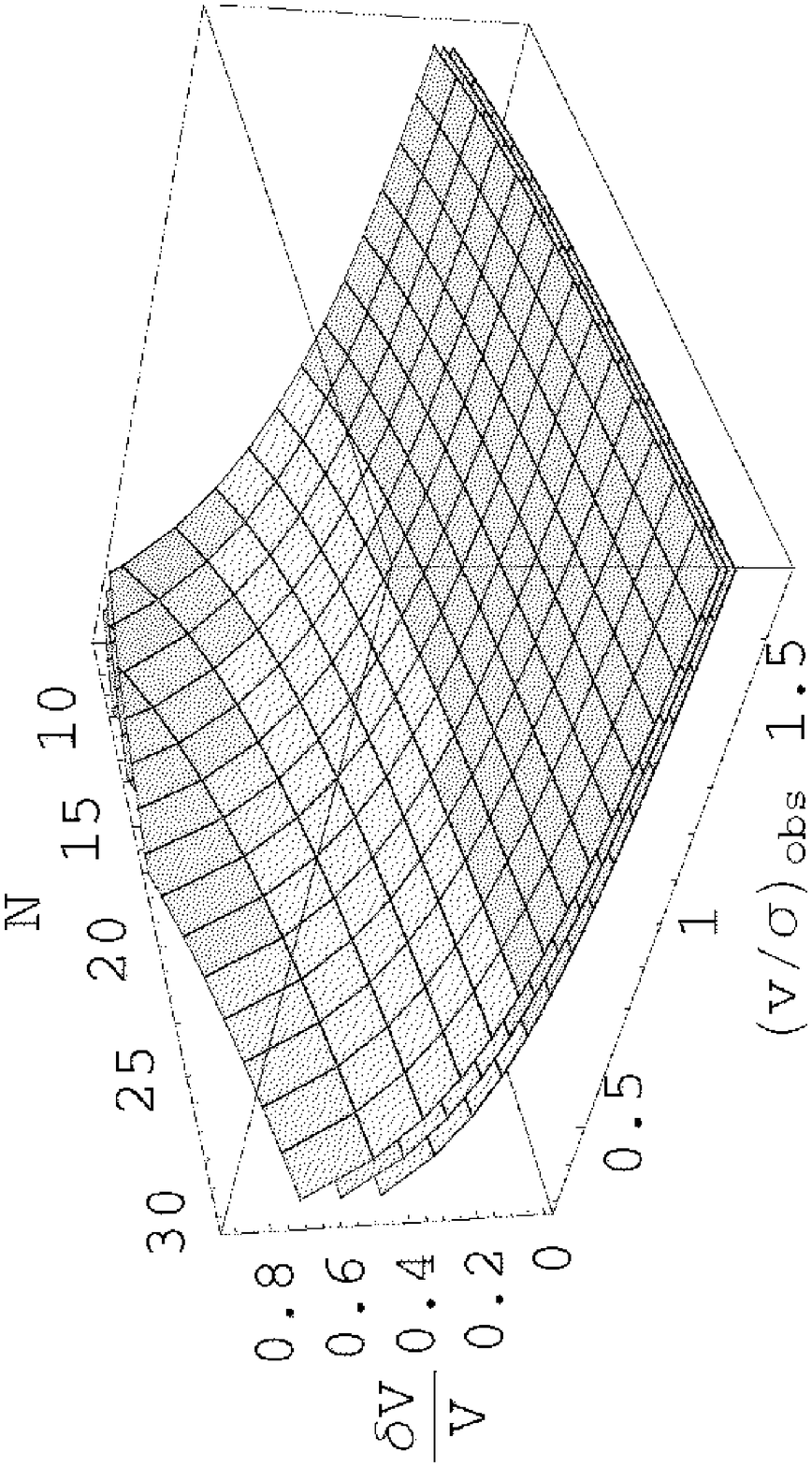,width=7.5cm,height=7.5cm,angle=-90}
\epsfig{figure=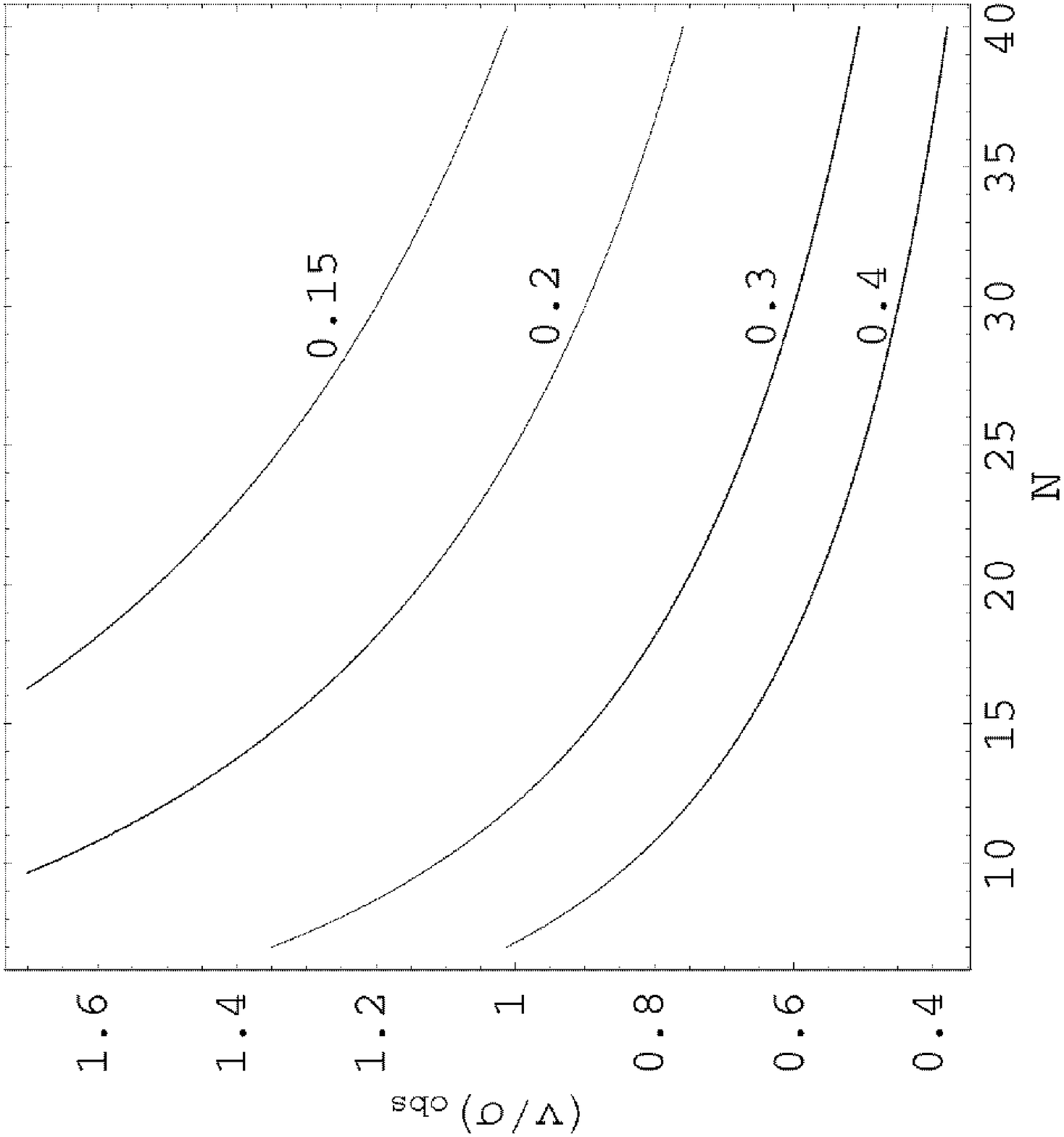,width=5.8cm,height=6.2cm,angle=-90}
\vspace{-1cm}
\caption{\footnotesize Precisions on rotation velocity. Left: behaviour of 
the generalised relative errors as function of $\overline{N}$ and 
$(v/\sigma)_\mathrm{obs}$, with the $\pm$SD surfaces which are computed  
from the errors on the $a$ and $b$-parameters. 
Right: 15\%, 20\%, 30\%, 40\% contours of the generalised relative errors.}
\label{univ}
\end{figure*}        
\noindent
\subsection{Errors on the rotation velocity}
We focus our analysis on the last spatial bin of the rotation velocity
curves, because the outer regions are those for which we wish to derive
the mass and angular momentum estimate. The statistical errors for the
velocity rotation measurements depend on the adopted model and bin
dimensions. The bin dimension determines the number statistics and
involves the velocity range over which we compute the average rotation
for a selected sample. The mass and rotation model determine the
dispersion profile which is needed for the equilibrium, and this
velocity dispersion enters in to the error budget of the rotation
measurement. 
In fact, using the definition of errors on the rotation velocity, we have
\begin{equation}
\frac{\delta V_\mathrm{rot}}{V_\mathrm{rot}}=
\frac{\sigma_\mathrm{p}}{\sqrt{\overline{N}}}\frac{1}{V_\mathrm{rot}}
\end{equation}
where $\sigma_\mathrm{p}=\sigma_\mathrm{V}\sqrt{1+\sigma_\mathrm{mea}^2/\sigma_\mathrm{V}^2}=C_\mathrm{V}\sigma_\mathrm{V}$ and
\begin{equation}
\frac{\delta V_\mathrm{rot}}{V_\mathrm{rot}}=
\frac{C_\mathrm{V}}{\sqrt{\overline{N}}}\frac{\sigma_\mathrm{V}}{V_\mathrm{rot}}=
\frac{C_\mathrm{V}}{\sqrt{\overline{N}}}\left(\frac{v}{\sigma}\right)^{-1}_\mathrm{obs}.
\label{eqerr}
\end{equation} 
\begin{table}
\caption[]{Here we list the set of parameter values for the fits displayed in Fig.~\ref{reldi} (right panel).}
\begin{tabular}{ccc}
\hline \multicolumn {3}{c}{\footnotesize{\bf Parameters in $\delta v/v$ fit}}\\
\hline
$\overline{N}$ & $A$ & $B$\\
\hline
29 & 0.185 $\pm$0.019 & 0.00$\pm$0.03\\
14 & 0.27 $\pm$0.05 & 0.00$\pm$0.07\\
13 & 0.28 $\pm$0.02 & 0.02$\pm$0.03\\
12.5 & 0.29 $\pm$0.02 & 0.015$\pm$0.020\\
 11 & 0.30  $\pm$0.02 & 0.05$\pm$0.03 \\
8.5 & 0.38  $\pm$0.03 & 0.03$\pm$0.04\\
 8 & 0.39  $\pm$0.04 & 0.02$\pm$0.05\\
 7 & 0.50 $\pm$0.04 & -0.02$\pm$0.06\\
\hline
\noalign{\smallskip}
\end{tabular}
\label{fits}
\end{table}
Here we want to check this dependence. For a fixed $\overline{N}$ (i.e. the 
mean number of particles in the last bin), one expects the 
relative errors to be reproduced by the following function
\begin{equation}
\frac{\delta v}{v}=A {(v/\sigma)_\mathrm{obs}}^{-1}+B.
\label{ervrcor}
\end{equation}

In Fig.~\ref{reldi} the best-fit for different $\overline{N}$, as in 
Table~\ref{fits} are shown. For these curves, B is always consistent 
with zero, while, from Eqs.~(\ref{eqerr}) and (\ref{ervrcor}), 
we expect A=$C_\mathrm{V}/\sqrt{\overline{N}}$ and $C_\mathrm{V}$=1.08,
for a typical $\sigma_\mathrm{mea}/\sigma_\mathrm{V}$=0.4 in the last bins. 
To test this dependence, the A parameter obtained for the different 
error profiles in Fig.~\ref{reldi} is interpolated with the function 
$a/\sqrt{\overline{N}}+b$; the best fit is for a=1.15$\pm$0.10 and 
b=-0.02$\pm$0.03, in agreement with the expected behavior. 

All these results were obtained for a wide range of models, so 
they can be generalised and used to estimate the
sample size needed to reach a targeted precision of the observed
kinematical quantities.

Once the kinematics is known from integrated light in the inner region
of a galaxy system, we can estimate the minimum sample needed
to reach a targeted precision in the outer regions, based on the inner 
$(v/\sigma_\mathrm{obs})$. 
In Fig.~\ref{univ} we show the dependence of 
rotation velocity precisions based on the results in Fig.~\ref{reldi}, 
as function of $(v/\sigma)_\mathrm{obs}$ and $\overline{N}$. 
In the same Figure, the contours for different precisions are shown.
The results of this analysis indicate that 1) the relative errors on the 
rotation velocity computed from a single measurement are consistent with 
those expected from the distribution of 100 simulation, 2) the 
precision on the rotational velocity is correlated with the internal 
kinematics of real systems and 3) this correlation is quite general 
and can be used in feasibility studies, once we need to estimate exposure 
times and plan observing proposals to reach a given sample size, i.e.
a targeted precision. 
For example, if one needs to reach an accuracy on $V_\mathrm{rot}$ of 20\%,
at least 30 PNe are needed in the last bin for systems were
$(v/\sigma > 1)$; to reach similar precision, larger samples are 
needed for systems with a lower value of $(v/\sigma)$.

\section{Summary and conclusions \label{conclu}}
We have developed an algorithm to simulate real measurements of discrete
radial velocity fields for a given mass model at equilibrium, using a
simplified DF in a factorized form, and with different sample sizes. 
We use this algorithm to address two important questions when using
discrete radial velocity fields as a tool to study the kinematics of 
the outer regions of elliptical galaxies:
\begin{enumerate}
\item Are the simple 3-parameter functional forms used to fit small 
sample of PNe radial velocities introducing biases on estimates of the 
kinematical quantities? 
\item How does the precision of a kinematical estimate depend on
the sample sizes, taking into account the rotation and inclination of the
system under study?
\end{enumerate}

\noindent
The answers are the following: \\
1) A bilinear fit to a sample coming from a system with a rotation 
curve like that of Centaurus A (Eq. \ref{rotin}) does introduce a bias
in the velocity dispersion profile at large radii. This problem is
important and pervades more sophisticated non-parametric analyses.
Any kind of smoothing algorithm applied to a small discrete sample of 
observed radial velocities, as in the study of NGC 1316 by Arnaboldi 
et al. (\cite{arn3}), is effectively fitting a more or less bilinear form
to the data. Our simulations show that this is bound to
introduce a bias in the estimate of the velocity dispersion profile
derived from the residual field, in particular for highly rotating 
systems. A flat rotation curve, or a rotation
curve profile derived from averaged data in bins along the line of
maximum gradient, should be adopted to derive the best estimate for
$\sigma$. Such a procedure was indeed used in the case of NGC 5128
(Hui et al. \cite{hui}). The overestimate in $\sigma$ caused by fitting a 
plane can be up to 20\% in $\sigma$, i.e. 40\% in the total mass. 
This bias correlates with the $(v/\sigma)$ ratio, in the sense that
higher rotation leads to stronger biases in $\sigma$. We also 
identified the most reliably estimated quantity derived when
using these simple 3-parameter fields: the line of maximum
gradient. This leads us to adopt the NFP, i.e. the analysis of the
binned quantities along this P.A., as the best approach free of
analysis-induced biases.\\
2) We have found that the precisions obtained on single measurements 
are consistent with those obtained from the simulated distributions.
Moreover we have found a generalised relation between the precision 
one can obtain for the observables in the outer regions as function of
sample size and inner kinematics. This empirical relation
can be very extensively used to plan and estimate observing times
to study the dynamics of the outer haloes of giant early type galaxies
with multi slit spectrographs like FORS2 or VIMOS on VLT,
or in slitless spectroscopy.
\begin{acknowledgements}
The authors are grateful to G. Busarello and O. Gerhard for their useful
comments and suggestions. The authors whish to thank M. Dopita for a careful reading of the manuscript before submission. N.R.N. is receiving financial support from the European Social Found.
\end{acknowledgements}

\section*{Appendix: the theoretical signal-noise ratio (SNR)}
\begin{figure}[h]
\centering
\epsfig{figure=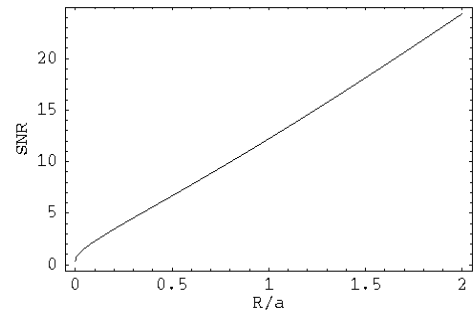,width=6.5cm,height=4.5cm}
\caption{\footnotesize The R-coordinate dependence of the $SNR$ for our
parameter choice.}
\label{SNRfig}
\end{figure}

\noindent
In our simulated velocity fields we introduce a cut-off in the projected
PN density distribution determined by a inner {\em limiting radius}, at 
which the incompleteness of the observed PNe sample due to the bright 
continuum light in the central part of an E galaxy becomes significant.
For a given luminosity profile, we consider all the PNe having magnitude 
$m=m_\mathrm{lim}$ which is the faintest magnitude in a PNLF for a complete sample, 
and we compute the SNR with respect to the galaxy background. We take a 
PN to be detected if it has a detected flux with SNR=9 (Ciardullo et al. 
\cite{ciar87}). Considering a negligible read-out-noise and sky background 
distribution, the SNR is defined as follows
\be
SNR=\frac{N_{5007}}{\sqrt{N_{5007}+N_\mathrm{gal}}}
\label{SNR}
\ee
where $N_{5007}$ are the photon counts detected for a single PN in the
5007 \AA\ [OIII] line and $N_\mathrm{gal}$ are the photon counts from the
continuum galaxy background\footnote{In Eq. (\ref{SNR}), we are considering the inner bright part of galaxies where the photon counts from the star continuum is much larger than the sky background, i.e. $N_\mathrm{gal}>>N_{sky}$. In the same equation, $N_{5007}>>N_{sky}$ by definition of detectability for a PN: SNR=9 implies that $N_{5007}>N_\mathrm{gal}>>N_{sky}$.}. These quantities are related to the incoming fluxes as follows
\be
N_{5007}=\frac{F_{5007}~t_\mathrm{esp}~S_\mathrm{tel}~\varepsilon}{E_{5007}}
\label{NPN}
\ee
where $F_{5007}$ is the flux received in the 5007 \AA\ line, $t_\mathrm{esp}$
is the exposure time, $S_\mathrm{tel}$ is the telescope surface, $\varepsilon$
is the total efficiency (instrumental efficiency + atmosphere
absorption), $E_{5007}$ is the energy of the photons at $\lambda=$5007
\AA 
\be
N_\mathrm{gal}=\frac{F_{V}}{E_{V}}\Delta \lambda~t_\mathrm{esp}~S_\mathrm{tel}~\varepsilon~S_\mathrm{sd}\simeq
\frac{F_{V}}{E_{5007}}\Delta \lambda~t_\mathrm{esp}~S_\mathrm{tel}~\varepsilon~S_\mathrm{sd}
\label{Ngal}
\ee
where $F_{V}$ is the flux per unit area received by the continuum galaxy 
background in the V-band, $\Delta \lambda$ is the pass band of a 
typical narrow filter, and $S_\mathrm{sd}$ is the area of the seeing disk
($S_\mathrm{sd}=\pi(3*FWHM/2.356)^2$).\\
Substituting (\ref{NPN}) and (\ref{Ngal}) in the relation (\ref{SNR}), we
find that
\be
SNR=F_{5007}\sqrt{\frac{t_\mathrm{esp}~S_\mathrm{tel}~\varepsilon}{E_{5007}
}}\frac{1}{\sqrt{F_{5007}+F_\mathrm{V}\Delta\lambda S_\mathrm{sd}}}
\label{SNRsos}
\ee
where
\be
F_\mathrm{V}=10^{-0.4(8.3268\left(\frac{R}{R_\mathrm{e}}\right)^{1/4}+12.7532+\mu_\mathrm{e}-(B-V))}\\
\ee
and
\be
F_{5007}=10^{-0.4(m_\mathrm{lim}+13.74)}
\ee
(Ciardullo et al. \cite{ciar89}). In Fig.~\ref{SNRfig}, we plot the $SNR$ for a 3.6m
telescope, 
using $T_\mathrm{esp}=3$ hrs, $\Delta \lambda=$60 \AA, FWHM=1.3$''$ and 
$\varepsilon$=0.5 and considering  $m_\mathrm{lim}$=27.2 mag, $\mu_\mathrm{e}$=21.24 mag arcsec$^{-2}$ and B-V=0.1 which are the typical values for an E galaxy in 
Virgo Cluster (Caon et al. \cite{caon}; McMillan et al. \cite{mcmil}). 
The condition $SNR(R_\mathrm{lim})=9$ implies $R_\mathrm{lim}=0.7 R_\mathrm{e}$.

\newpage

\newpage
\begin{figure*}
\vspace{-0.7cm}
\centering
\epsfig{figure=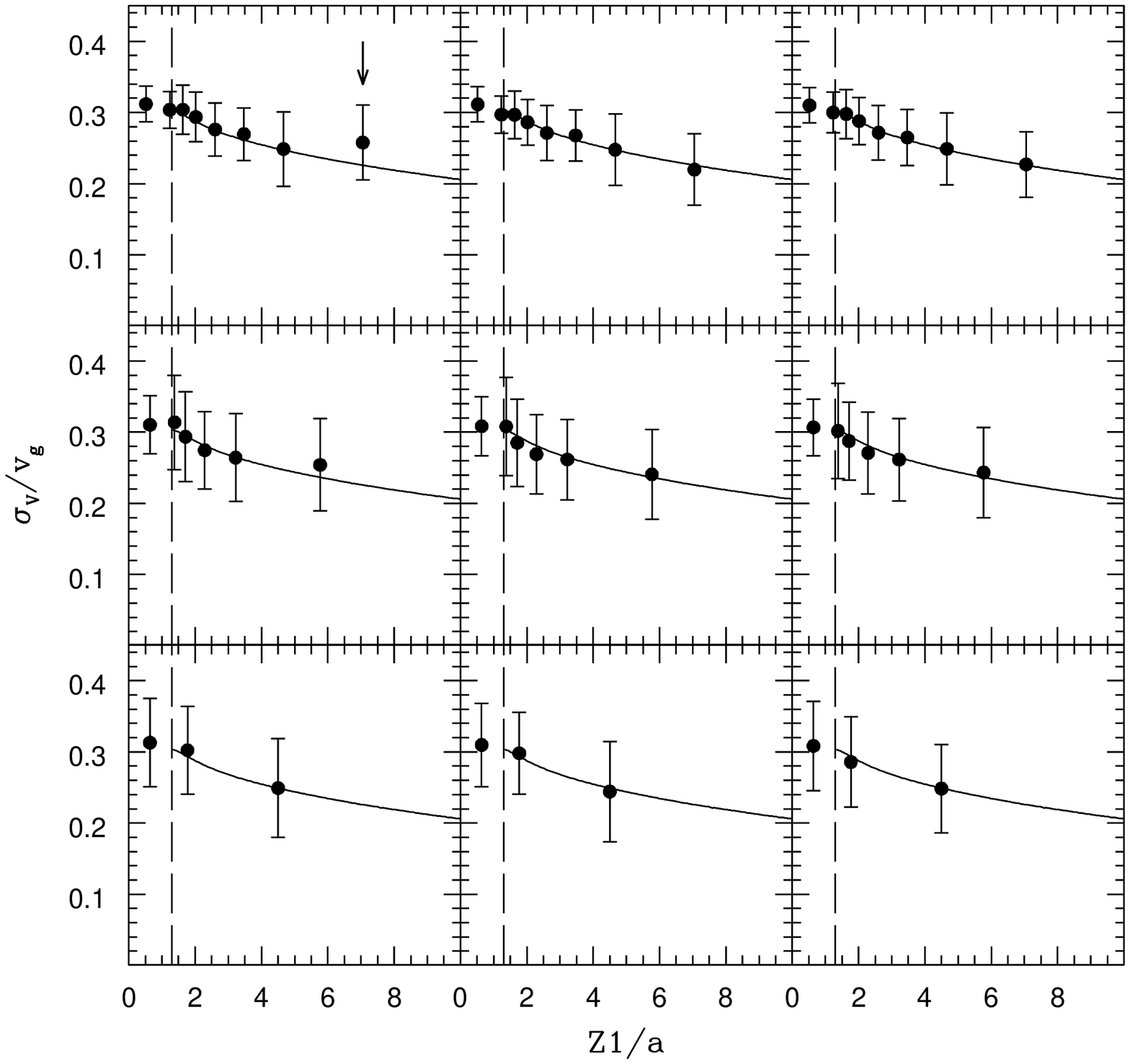,width=12cm,height=12cm}
\vspace{-0.7cm}
\vspace{-0.5cm}
\centering
\epsfig{figure=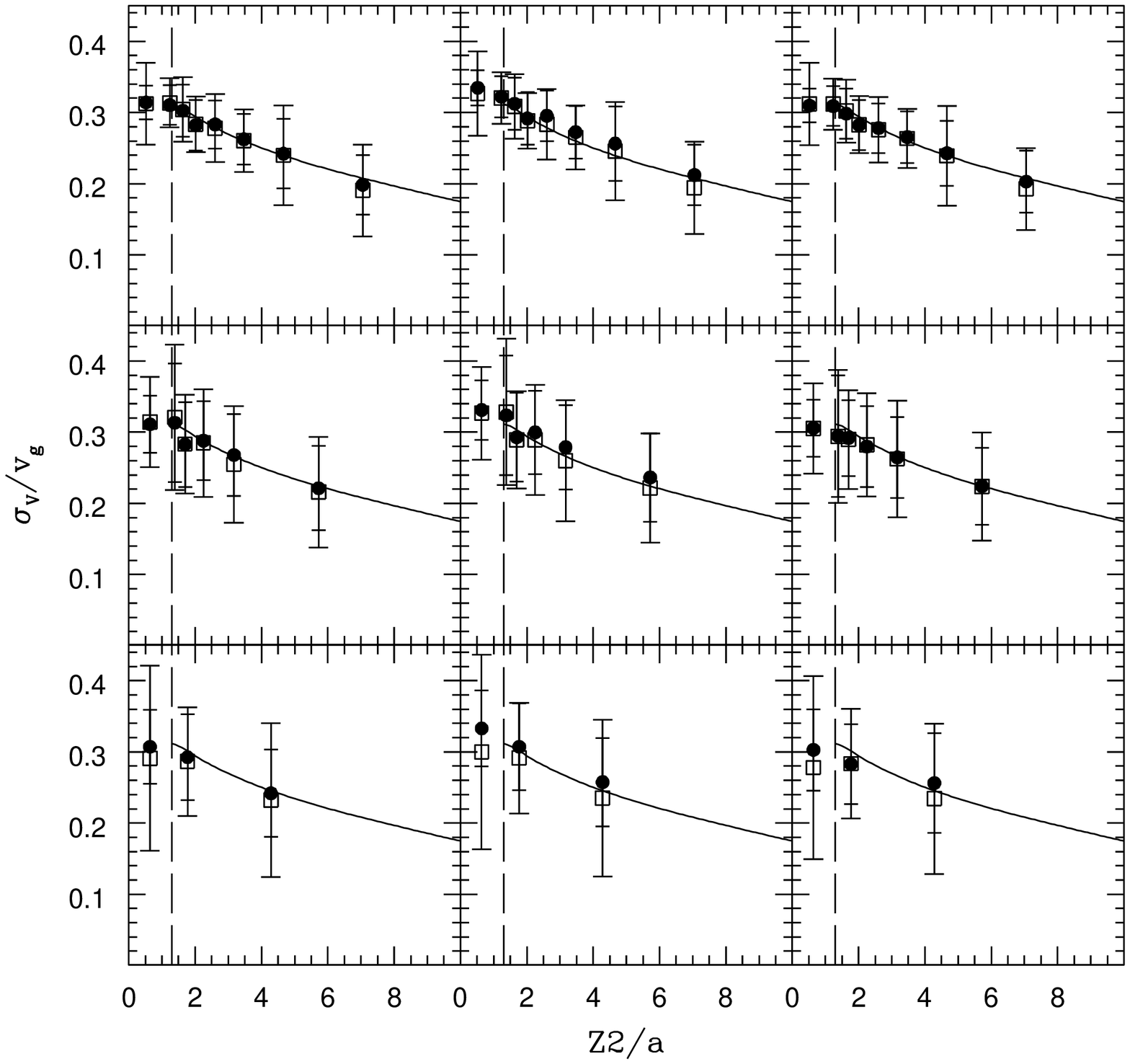,width=12cm,height=12cm}
\vspace{-0.5cm}
\caption{\footnotesize Velocity dispersion profiles for rotating model with DM along Z1 (upper panel, $V_\mathrm{max}=0.30v_\mathrm{g}$) and Z2 (lower panel, $V_\mathrm{max}=0.23v_\mathrm{g}$): {\em edge-on case}. Full points
are the mean values obtained on 100 simulations, solid line is the expected
velocity dispersion profile, empty squares are the estimates using $dZ_\mathrm{min}$. The bias introduced by the BF is indicated by the arrow (upper panel 500 PNe case; see also Fig.~\ref{bias}). In each panel:
{\bf
Top}: 500 PNe; {\bf
middle}: 150 PNe;{\bf bottom}: 50 PNe. {\bf Left}: bilinear fit procedure;
{\bf
center}: no-fit procedure; {\bf right}: flat-curve procedure.} 
\label{self_cyl20}
\end{figure*}

\begin{figure*}
\vspace{-0.7cm}
\centering
\epsfig{figure=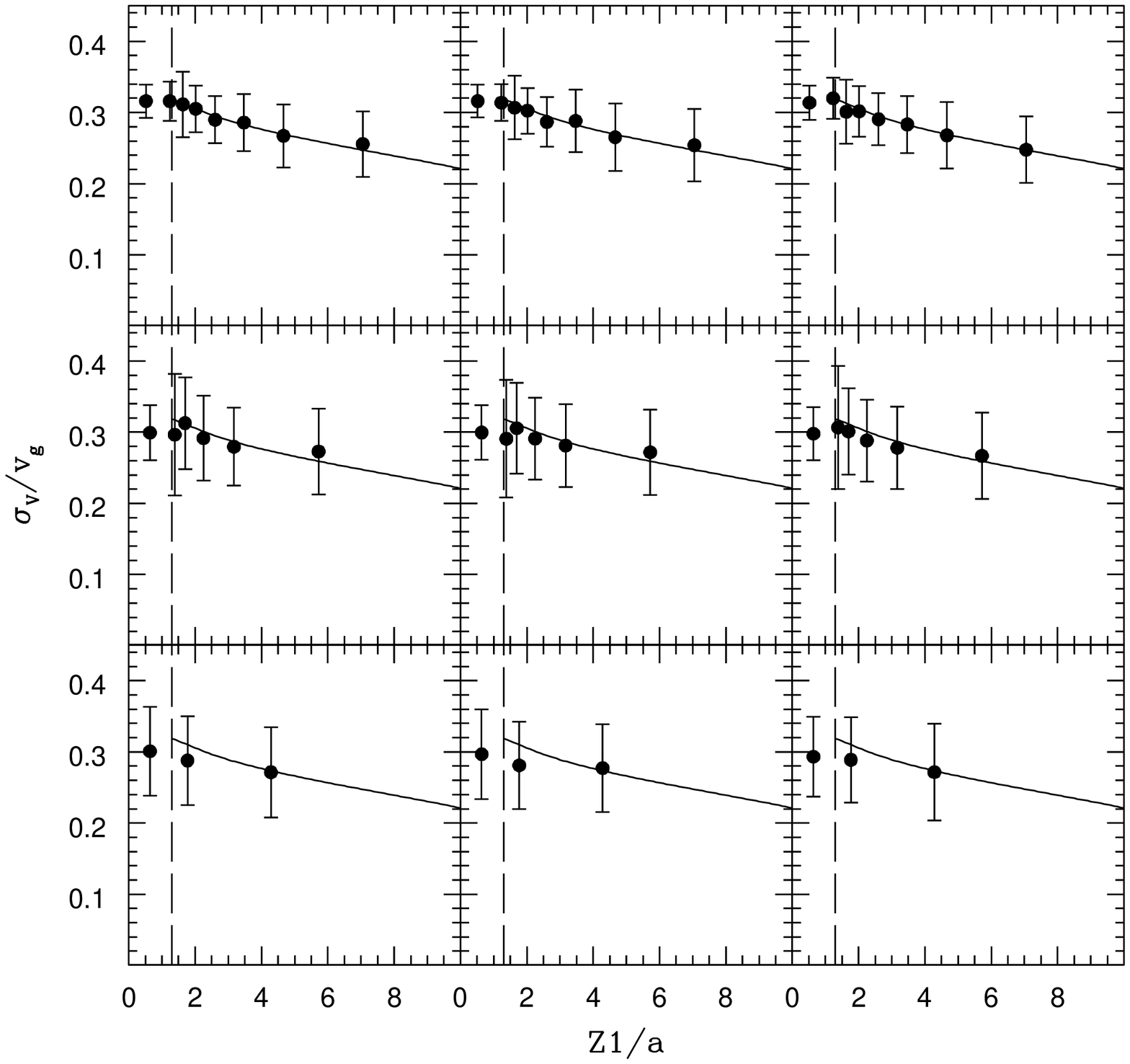,width=12cm,height=12cm}
\vspace{-0.7cm}
\vspace{-0.5cm}
\centering
\epsfig{figure=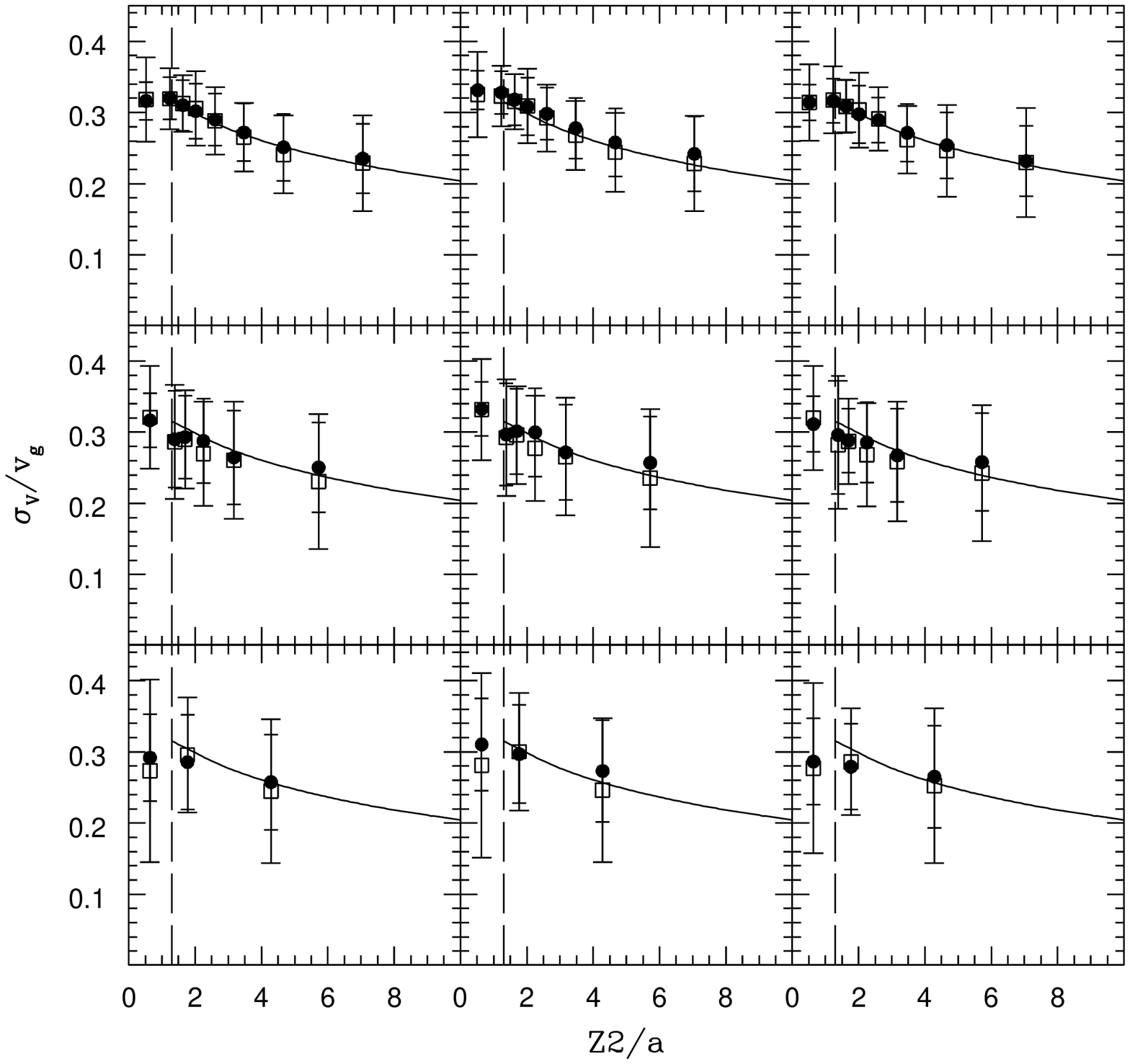,width=12cm,height=12cm}
\vspace{-0.5cm}
\caption{ \footnotesize  Velocity dispersion profiles for rotating model with DM along Z1 (upper panel, $V_\mathrm{max}=0.30v_\mathrm{g}$) and Z2 (lower panel, $V_\mathrm{max}=0.23v_\mathrm{g}$). LOS: -45:-45:-45. Symbols have the same meaning as in Fig.~\ref{self_cyl20}. The amount of the biases is negligible here because the $(v/\sigma)_\mathrm{obs}=V_\mathrm{rot}/\sigma_\mathrm{V}$ ratio is lower. In each panel:
{\bf
Top}: 500 PNe; {\bf
middle}: 150 PNe;{\bf bottom}: 50 PNe. {\bf Left}: bilinear fit procedure;
{\bf
center}: no-fit procedure; {\bf right}: flat-curve procedure.} 
\label{self_cyl245}
\end{figure*}

\end{document}